\documentclass[aps,prl,reprint,superscriptaddress]{revtex4-2}
\usepackage{graphicx}
\usepackage{amsmath}
\usepackage{amssymb}
\usepackage{mathrsfs}
\usepackage[T1]{fontenc}
\usepackage{calligra}
\usepackage{array}
\usepackage{xcolor}
\usepackage{float}
\usepackage{soul}
\usepackage{bm}
\usepackage[toc,page]{appendix}
\usepackage{braket}
\usepackage{wrapfig}
\usepackage[colorlinks=true, linkcolor=blue, urlcolor=blue,citecolor=blue]{hyperref}

\hypersetup{colorlinks,linkcolor=blue,citecolor=blue,urlcolor=blue}

\begin{document}
	\title{Ultrafast all-electrical universal nano-qubits}
        
	\author{David T. S. Perkins}
	\affiliation{School of Physics, Engineering and Technology and York Centre for Quantum Technologies, University of York, YO10 5DD, York, United Kingdom}
	
	\author{Aires Ferreira}
	\email[]{aires.ferreira@york.ac.uk}
	\affiliation{School of Physics, Engineering and Technology and York Centre for Quantum Technologies, University of York, YO10 5DD, York, United Kingdom}
	
    
    \begin{abstract}
        We propose how to create, control, and read-out real-space localised spin qubits in proximitized finite graphene nanoribbon (GNR) systems using purely electrical methods. Our proposed \textit{nano-qubits} are formed of in-gap singlet-triplet states that emerge through the interplay of Coulomb and relativistic spin-dependent interactions in GNRs placed on a magnetic substrate. Application of an electric field perpendicular to the GNR heterostructure leads to a sudden change in the proximity couplings, i.e. a quantum quench, which enables us to deterministically rotate the nano-qubit to any arbitrary point on the Bloch sphere. We predict these spin qubits to undergo Rabi oscillations with optimal visibility and frequencies in excess of 10 GHz. Our findings open up an avenue for the realisation of graphene-based quantum computing with ultra-fast all-electrical methods.
    \end{abstract}
    
    \maketitle
	
    
    Quantum computing is a new technological frontier with paradigm-shifting capabilities in fields as diverse as quantum chemistry, cyber security, and machine learning \cite{Nielsen_book,Rev_QC_chemistry_19,Rev_QC_machine_learning_17}. 
    Semiconductor platforms for spin-based quantum information processing are a promising path towards realising stable qubits. Amongst the most prominent spin qubit host systems are quantum dots (QDs) and donors \cite{Review_Harvey_22}, with group-IV semiconductors being at the forefront of current efforts in the field due to their long spin coherence times and scalability potential \cite{Pla2012,Laird2013,Hendrickx2021,Park2023}. Leveraging these efforts, recent experiments with silicon qubits have demonstrated one- and two-qubit gate operations yielding fidelities exceeding the thresholds of leading quantum error-correcting codes \cite{ErrorC_2022_1,ErrorC_2022_2,Madzik2022}, taking spin qubits one step closer to fault-tolerant quantum computing \cite{Fowler_12}.
    
    However, our ability to operate electron spin qubits has predominantly relied upon external magnetic fields to lift the spin degeneracy of the electronic states, which has hindered device miniaturisation and set fundamental limits on qubit manipulation speeds \cite{Review_Vandersypen_17}. Two methods are routinely implemented to rotate spin qubits: electron spin resonance (ESR) and electric-dipole spin resonance (EDSR). The former requires the use of an oscillating magnetic field to drive resonant transitions between the different spin states of a qubit   \cite{Koppens2006,Vahapoglu2021}. In contrast, EDSR can only be used in systems possessing a spin-orbit field \cite{Corna2018,Borjans2019,Hosseinkhani2022,Wang2022}, where a constant magnetic field in combination with an oscillating electric field can encourage dipole transitions between the qubit's states, thus simplifying  device architectures by allowing for electrically-driven qubits  \cite{Nowack2007,Leon2020}. To date, the fastest EDSR-driven qubits have been observed in a germanium hut wire displaying coherent Rabi oscillations at a rate of 540 MHz with a 0.1 T field \cite{Wang2022}. Removing the need for external magnetic fields is a critical factor in breaking the gigahertz (GHz) barrier and improving the energy efficiency of solid-state quantum computers, yet it remains an open question.

     \begin{figure*}
        \centering
        \includegraphics[width=0.7\textwidth]{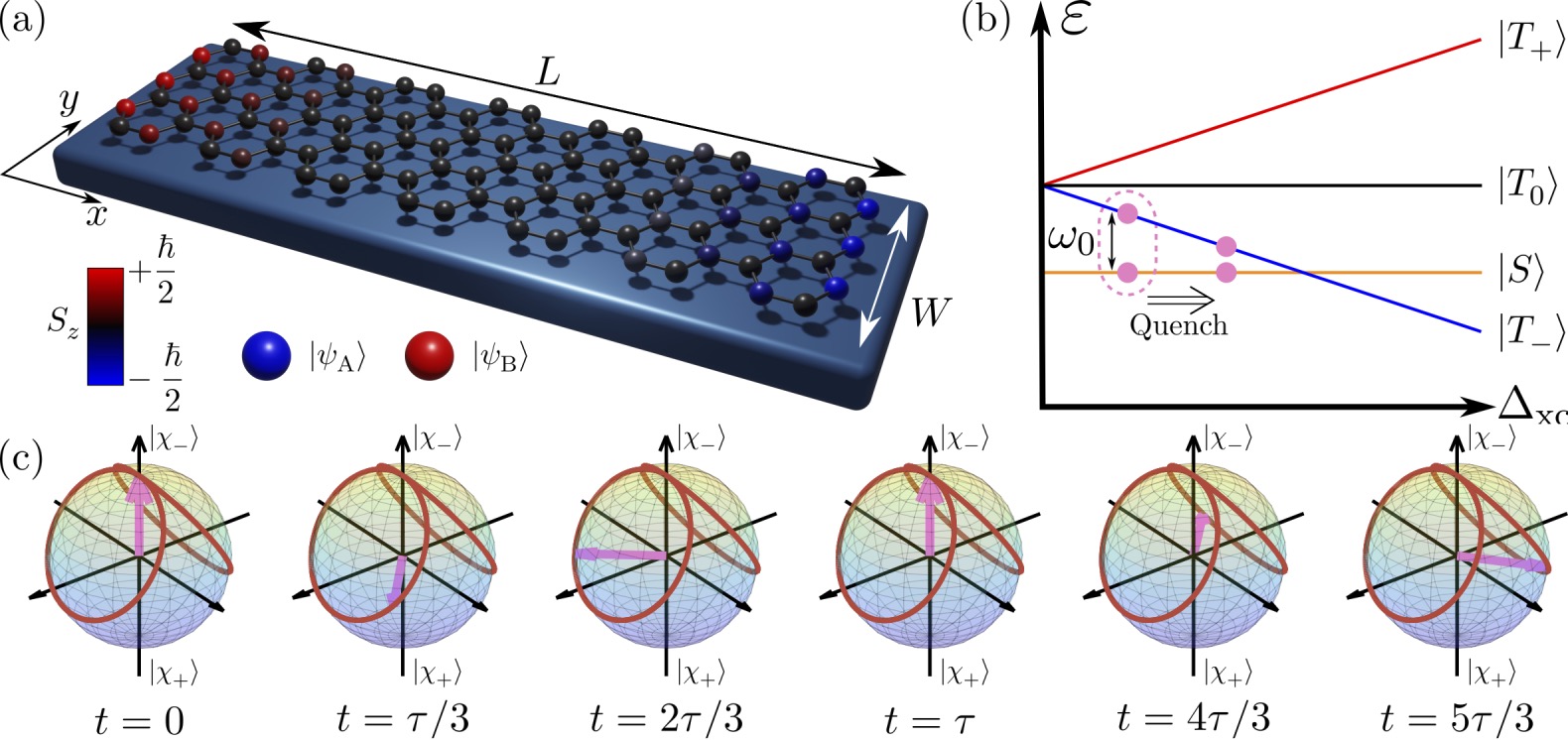}
        \caption{(a): Schematic of an $L \times W$ atom GNR placed on a magnetic substrate with sublattice localised states appearing at opposite ends of the GNR. (b): Schematic of how the singlet and triplet energy levels evolve with $\Delta_{\text{xc}}$. The qubit will be formed from the $\ket{S}$ and $\ket{T_{-}^{\null}}$ states, with a gap of $\omega_{0}$ between them, and is denoted by the pink dashed box. Applying a quench that suddenly increases $\Delta_{\text{xc}}$ shifts the energies of the singlet and triplet, triggering the spin qubit dynamics. (c): An example orbit traversed by the qubit (pink arrow) around the Bloch sphere post quench, with $\tau = 1/f_{\text{R}}$ being the Rabi oscillation period. The $\ket{\chi_{\pm}^{\null}}$ are the ground state ($-$) and excited state ($+$) of the qubit resulting from superpositions of $\ket{S}$ and $\ket{T_{-}^{\null}}$.}
        \label{GNR_energy_levels}
    \end{figure*}
    
    The advent of graphene has inspired alternative routes to creating and manipulating quantum bits courtesy of its low dimensionality and exotic Dirac spectrum \cite{Trauzettel2007}. Recent studies have had great success in the electrostatic confinement of single electrons in monolayer and bilayer graphene \cite{Eich2018,Banszerus2020,Kurzmann2021,Garreis2021}. The ability to electrically tune the band gap in bilayer graphene combined with its magnetically-addressable valley pseudospin suggests interesting avenues for quantum computing akin to spin-valley qubit operation in silicon \cite{Collard_19,Banzerus_21,Jock_22}. Prospects for developing bona fide qubits in bilayer graphene QDs have been boosted with reports of spin lifetimes exceeding 0.2 ms in a 1.9 T magnetic field \cite{Banszerus2022} and single-shot spin readout \cite{Gachter2022}. Despite these advances,  technical challenges in nanofabrication will need to be overcome to operate qubits in the low magnetic field regime, where spin-valley coherence is maximized \cite{Banszerus_23}. More recently, significant advancements in atom-by-atom fabrication of molecular nanographenes have been made that allow for the creation of designer finite graphene nanoribbons (GNRs) with precise shapes and edge morphology \cite{Slota2018,Chen2020,Abbassi2020,Niu2023}. Due to their molecular precision, these auspicious GNRs boast  spin relaxation times on the order of milliseconds at temperatures as large as 10 K  \cite{Slota2018}, offering exciting prospects for further investigations. A particularly attractive possibility is to encode a logical qubit in sublattice-split states of a finite GNR that are inherently real-space localised, thus providing complementary attributes to the delocalised spin-valley qubits in gated-defined QDs. Atomically precise GNRs may therefore establish the next epoch of quantum computing for qubit encoding via natural quantum confinement.
    
    In this Letter, we propose a type of universal spin qubit that can be manipulated via purely electrical methods (i.e. a universal all-electrical nano-qubit), encoded in pairs of quasi-degenerate, real-space localised in-gap states that are ubiquitous in zig-zag GNRs, GNRs with tunable extensions, and molecular nanographenes \cite{Fujita1996,Nakada1996,Ortiz2018,Ortiz2019,Pizzochero2021}. Prototypical real-space localised  states, such as edge states, emerge in sublattice-compensated nanographenes and possess well-defined spin and sublattice-pseudospin, stabilised via antiferromagnetic correlations \cite{Son2006,Rossier07}. Our scheme utilises the vast potential of proximity effects to create bespoke nanostructures with optimal relativistic electronic structure realized through graphene's pairing with other two-dimensional (2D) materials \cite{Offidani2017,Island2019,Sierra2021}. Specifically, atomically thin magnets (e.g. Cr${}_{2}$Sn${}_{2}$Te${}_{6}$) provide an intrinsic time-reversal symmetry breaking mechanism, by which to lift the spin triplet degeneracy, and induce proximity magnetic exchange coupling (MEC) up to 6.8 meV in graphene \cite{Zhang2015,Wu2021,Zollner2020,Zollner2022}. Furthermore, atomically thin semiconductors in the group-VI dichalcogenide family have been shown to induce Rashba spin-orbit coupling (SOC) as large as 13 meV in graphene \cite{Wang2019,Rao2023}. Therefore, proximity effects can be exploited to gain all-electrical access to qubit encoding spaces of a naturally confined GNR. What makes our system stand out, as we show below, is that  electrical driving of qubit rotation can be achieved using an applied out-of-plane electric field to rapidly change the  Rashba SOC and MEC induced by a partner material. Such a quantum quench protocol provides deterministic control akin to the ESR and EDSR approaches, without the need for external magnetic fields. The tunability of proximity effects in graphene nanostructures is shown to easily allow for coherent Rabi oscillations with perfect visibility and frequencies far exceeding the 1 GHz barrier, approaching the terahertz (THz) regime in small GNRs. Such nano-qubits are thus predicted to outperform even the fastest group-IV semiconductor-based spin-orbit qubits, with current record Rabi frequencies of 540 MHz in Ge \cite{Wang2022}. We furthermore show that the nano-qubits can be detected in a graphene bridge setup similar to Ref. \cite{Niu2023} by using a single-shot read-out protocol akin to Refs. \cite{Gachter2022,Elzerman2004}, wherein a charge detector in the form of a secondary QD is used to detect the electrostatic changes due to the loading/unloading of an electron from the proximity-coupled GNR. Graphene nanostructures of this scale have been fabricated using bottom-up methods in a series of recent studies \cite{Slota2018,Chen2020,Abbassi2020,Niu2023}, wherein long spin lifetimes ($T_{1} \sim 5$ ms and $T_{2} \sim 0.4 \, \mu$s at 2 K) were observed which may benefit their ability to host nano-qubits.
    
    
    \textit{\textit{Model}.---}   We consider the class of particle-hole symmetric finite GNR systems whose spinless non-interacting Hamiltonians exhibit a pair of in-gap states \cite{Ortiz2018,Ortiz2019,Pizzochero2021,Tepliakov2023} (i.e. quasi-degenerate zero energy states), $\ket{\psi_{\pm}}$, lying close to and symmetrically about zero energy at $\varepsilon_{\pm}$, and that are energetically well separated from all other states. An example of such a system is depicted in Fig. \ref{GNR_energy_levels}a, which we use as proxy for the quality of the envisaged nano-qubits. From these low-energy states, we can construct two sublattice-localised states, $\ket{\psi_{\text{A(B)}}} = (\ket{\psi_{+}} \pm \ket{\psi_{-}})/\sqrt{2}$ \cite{Ortiz2018}, which are further localised at the opposite zig-zag edges of the GNR (see Fig. \ref{GNR_energy_levels}a). In what follows, we refer to these states as quasi-zero energy modes (QZEMs). The effective two-site Hubbard Hamiltonian \cite{Ortiz2019,Supplementary_material} for a proximitized GNR in terms of the QZEMs is given by
    \begin{equation}
    \begin{split}
        \widetilde{H} = &\sum_{\sigma} (\tilde{t} \, a_{\sigma}^{\dagger}b_{\sigma}^{\null} + \tilde{\lambda}_{\text{R},\sigma} \, a_{\sigma}^{\dagger}b_{\bar{\sigma}}^{\null} + h.c.) \\
        &+ \sum_{\nu = a,b} \Big[ \Delta_{\text{xc}} (\nu_{\uparrow}^{\dagger}\nu_{\uparrow}^{\null}-\nu_{\downarrow}^{\dagger}\nu_{\downarrow}^{\null}) + \widetilde{U}(\nu_{\uparrow}^{\dagger}\nu_{\uparrow}^{\null} \nu_{\downarrow}^{\dagger}\nu_{\downarrow}^{\null}) \Big],
        \label{Hubbard_Dimer_Hamiltonian}
    \end{split}
    \end{equation}
    where $\tilde{t} = (\varepsilon_{+}-\varepsilon_{-})/2 = \delta/2$ is the QZEM hybridisation energy, $\tilde{\lambda}_{\text{R},\sigma} = \text{sgn}(\sigma) \tilde{\lambda}_{\text{R}}$ is the effective Rashba SOC strength due to interfacial breaking of mirror reflection symmetry, $\Delta_{\text{xc}}$ is the proximity-induced MEC, $\widetilde{U}$ is the effective on-site Hubbard interaction, $\bar{\sigma} = -\sigma$, and $a_{\sigma}^{(\dagger)}$ and $b_{\sigma}^{(\dagger)}$ annihilate (create) a particle with spin $\sigma$ in state $\ket{\psi_{A}}$ and $\ket{\psi_{B}}$, respectively. For half-filling, the two-site model leads to a six-dimensional Fock space spanned by the basis $\{\ket{2,0},\ket{0,2},\ket{\uparrow,\uparrow},\ket{\downarrow,\downarrow},\ket{\downarrow,\uparrow},\ket{\uparrow,\downarrow}\}$, where the first and second states represent doubly occupied A and B sites, respectively, whilst the remainder of the states correspond to the different spin configurations of a single occupation on each site. Recasting $\widetilde{H}$ in this basis, we find that the lowest two eigenstates can be isolated by tuning the MEC, see Fig. \ref{GNR_energy_levels}b. In the limit of vanishing Rashba SOC, these low-lying states can be identified as the spin singlet, $\ket{S}$, and spin triplet, $\ket{T_{-}}$, that entangle the spin and sublattice-pseudospin degrees of freedom. It is these states that shall form the nano-qubit. Having a sizable Rashba SOC is crucial to enable resonant transitions between the singlet and triplet. The use of a Rashba SOC generated by an STM tip   to manipulate spins in nanographenes via EDSR was discussed in Ref. \cite{Ortiz2018}. However, there are two notable differences between the nanoqubit proposed here and previous work. First, the use of proximity-induced SOC yields extremely large values of Rashba SOC (on the order of $10$ meV), which grants easy electrical access to optimal Rabi oscillations through modulation of the proximity couplings via an electric field, as shown below. Second, degeneracies in the qubit-operating manifold are lifted due to the breaking of time-reversal symmetry by MEC, therefore entirely removing the need for external magnetic fields.
    
    Isolating the nano-qubit to first order in the Rashba coupling yields the effective qubit Hamiltonian:
    \begin{equation}
        \mathcal{H} = -\frac{\omega_{0}}{2} \sigma_{z} + g_{\text{R}} \sigma_{x},
        \label{Qubit_Hamiltonian}
    \end{equation}
    where $\sigma_{x,z}$ are Pauli matrices, $\omega_{0} = -\Delta_{\text{xc}} + (\mathcal{S} - \widetilde{U})/2$, $g_{\text{R}} = 8\tilde{t}\tilde{\lambda}_{\text{R}}[\widetilde{U}+\mathcal{S}]^{-1}[2+32\tilde{t}^{2}/(\widetilde{U}+\mathcal{S})^{2}]^{-1/2}$, and $\mathcal{S} = \sqrt{16\tilde{t}^{2}+\widetilde{U}^{2}}$. From Eq. \ref{Qubit_Hamiltonian} we can immediately see that the role played by the symmetry-breaking Rashba SOC is to allow transitions between the singlet and triplet manifolds. The qubit manifold is separated from the triplet states $T_{+}$ and $T_{0}$ by max$[|\Delta_{\text{xc}}|$, $|\mathcal{S}-\widetilde{U}|/2]$, depending on whether $\ket{S}$ or $\ket{T_{-}}$ is the ground state.
    
    \begin{figure}
        \centering
        \includegraphics[width=0.7\linewidth]{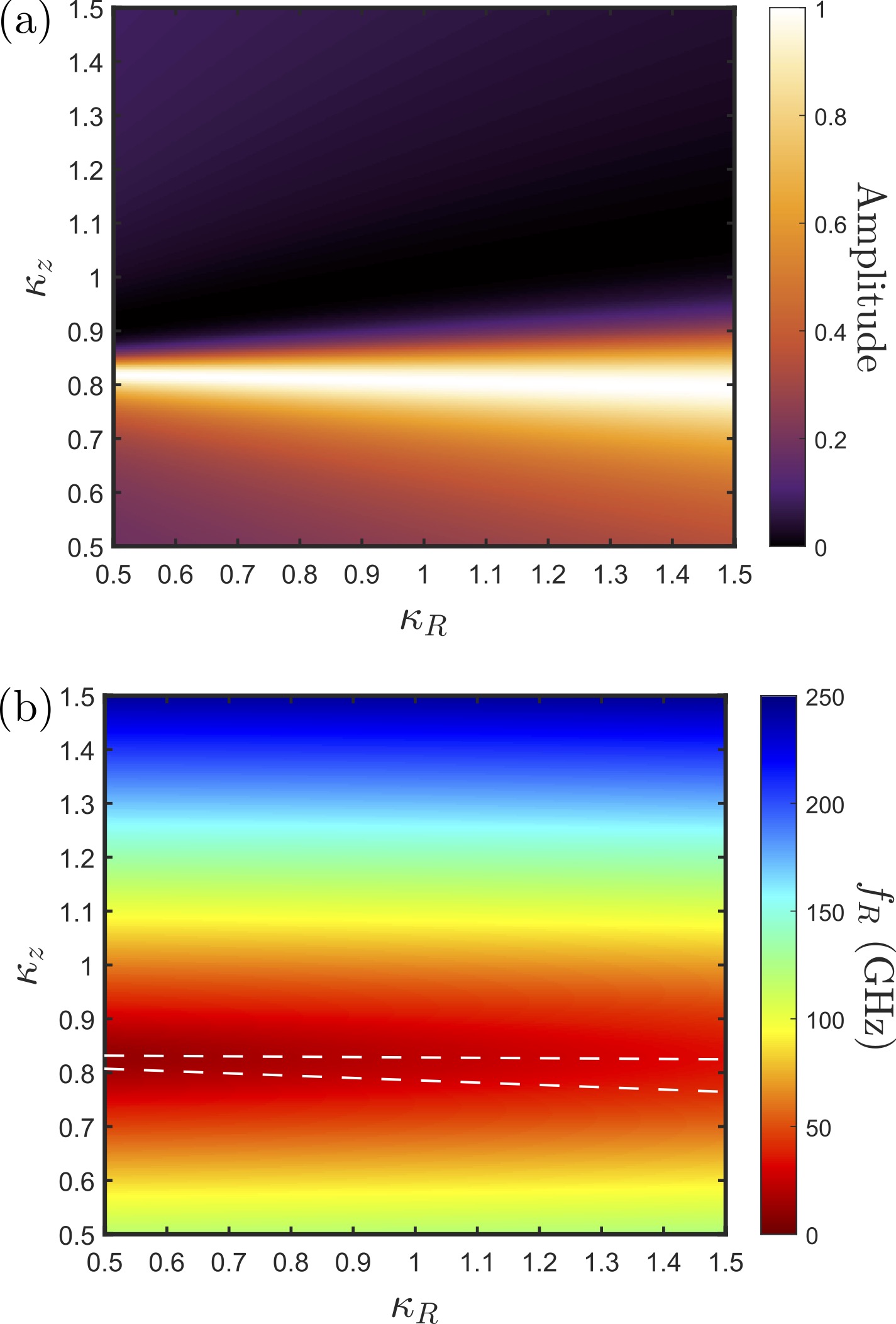}
        \caption{Resulting Rabi oscillation amplitude (a) and frequency (b) due to a quantum quench in a 36$\times$7 atom GNR. The Rabi oscillations can be seen to have frequencies in the range of 1 to 100 GHz for small quenches. The white dashed line in (b) highlights the region of near 100\% Rabi amplitudes from (a). The GNR parameters we have used here are: nearest-neighbour hopping energy $t = 2.7$ eV \cite{Castro2009}, on-site Hubbard repulsion $U = 2t$ \cite{Feldner2010,Feldner2011,Wehling2011,Jung2011,Goli2016}, Rashba coupling $\lambda_{\text{R}}^{(1)} = 10$ meV and $\Delta_{\text{xc}}^{(1)} = 1.5$ meV. Note that these are bare values characteristic of the entire GNR and are not the effective parameters appearing in Eq. \ref{Hubbard_Dimer_Hamiltonian}. They yield gaps of $\delta = 18$ meV and $\Delta = 0.28$ meV with a clear separation of $\ket{S}$ and $\ket{T_{-}}$ from the other four states. The corresponding effective parameters are: $\tilde{\lambda}_{\text{R}} = -0.49$ meV, $\tilde{t} = 9.1$ meV, and $\widetilde{U} = 0.26$ eV. The relations between the effective parameters and the real characteristic parameters can be found in Ref. \cite{Supplementary_material}.}
        \label{Quench_plots}
    \end{figure}
    
    
    \textit{\textit{Universal Qubit Control}.---} The electrical manipulation of single electron spins is traditionally achieved via EDSR \cite{Corna2018,Borjans2019,Hosseinkhani2022,Wang2022,Rashba1960}, with recent works having realized fast qubits in silicon nanowire quantum dots \cite{Corna2018} and hole qubits in Ge \cite{Wang2022}, as well as predicting Rabi frequencies up to 250 MHz in 2D semiconductors \cite{Brooks2020}. Here we put forward an alternative approach, a quantum quench, that takes advantage of the superior electrical tunability of atomically thin systems \cite{Yang2016,Shcherbakov2021,Amann2022} to realize all-electrical qubits. By applying a pulse gate voltage across the GNR heterostructure, the effective qubit parameters in Eq. \ref{Qubit_Hamiltonian} can be efficiently modulated above or below their zero-field values. The resulting change in the Rashba SOC and MEC is sensitive to the choice of materials used to proximitize the GNR, their structural orientation, and the direction of applied field, with variations as large as 10--30\% in several candidate materials, such as Cr${}_{2}$Ge${}_{2}$Te${}_{6}$ \cite{Gmitra2015,Zollner2022}. This fine degree of electrical tunability offers a wide range of possible control quenches achievable within state-of-the-art GNR experimental platforms.
    
    Let us start by considering a system whose MEC and effective Rashba coupling are given by $\Delta_{\text{xc}}^{(1)}$ and $\tilde{\lambda}_{\text{R}}^{(1)}$, respectively, for all time $t<0$. The corresponding qubit will then be governed by the Hamiltonian in Eq. \ref{Qubit_Hamiltonian} with $\omega_{0}^{\null} = \omega_{0}^{(1)}$ and $g_{\text{R}}^{\null} = g_{\text{R}}^{(1)}$. We denote the excited state and ground state of this qubit by $\ket{\chi_{\pm}^{(1)}}$, respectively. At $t = 0$ we apply the quantum quench to rapidly change the MEC and SOC, which in turn changes the parameters of the qubit Hamiltonian, $\omega_{0}^{(1)} \rightarrow \omega_{0}^{(2)}$ and $g_{\text{R}}^{(1)} \rightarrow g_{\text{R}}^{(2)}$. Prior to the quench the qubit will be in the ground state $\ket{\chi_{-}^{(1)}}$. After the quench, the qubit will be in the time evolving state $\ket{\psi(t)} = e^{-i\mathcal{H}_{2}t} \ket{\chi_{-}^{(1)}}$, and we find the probability of measuring the qubit in state $\ket{\chi_{+}^{(1)}}$ to be
    \begin{equation}
        P(t) = \sin^{2}\left(\frac{\Delta_{2}t}{2\hbar}\right) \sin^{2}(2\psi_{12}),
        \label{Qubit_probability}
    \end{equation}
    where $\Delta_{2} = E_{+}^{(2)} - E_{-}^{(2)}$ is the difference in energies of the eigenstates for the post-quench system, $\psi_{12} = \psi_{1} - \psi_{2}$ is the singlet-triplet quench mixing, and $\psi_{1(2)}$ are the pre quench (post quench) singlet-triplet angles of the eigenstates; see Ref. \cite{Supplementary_material} for details on their parameter dependence. From Eq. \ref{Qubit_probability}, the Rabi frequency for a quantum quench is readily seen to be $f_{\text{R}} = \Delta_{2}/h$, while the amplitude of the Rabi oscillations is set by $\psi_{12}$.
    
    To explore the ranges of Rabi frequencies and amplitudes that can be achieved in such GNRs, let us parameterise our quench in terms of $\kappa_{\text{R}}$ and $\kappa_{z}$, where $\tilde{\lambda}_{R}^{(2)} = \kappa_{\text{R}} \tilde{\lambda}_{\text{R}}^{(1)}$ and $\Delta_{\text{xc}}^{(2)} = \kappa_{z} \Delta_{\text{xc}}^{(1)}$. The range of possible Rabi frequencies and amplitudes are shown in Fig. \ref{Quench_plots} for a 36$\times$7 atom GNR. We see that a decrease of 15 to 20\% in $\Delta_{\text{xc}}$ yields the largest oscillations with amplitudes close to 100\%. Interestingly, a change in the Rashba coupling has drastically less impact on the qubit dynamics in comparison to the exchange interaction (see Fig. \ref{Quench_plots} and Ref. \cite{Supplementary_material}). This is ideal from a practical perspective as it makes our qubit control scheme an effective one-dimensional problem; the MEC quench acts as the primary governing parameter of the qubit evolution in parameter space. Focusing on the region of high amplitudes, we find that the associated Rabi frequencies are in excess of 10 GHz; note that the fastest reported qubit reported to date was observed in Ge quantum dots with a Rabi frequency of 540 MHz \cite{Wang2022}. As an example, let us consider a quench with $\kappa_{\text{R}} = 1.1$ and $\kappa_{z} = 0.8$. In this case we obtain Rabi oscillations with an amplitude of 99.9\% and a frequency of 27 GHz. Furthermore, for the choice of parameters listed in Fig. \ref{Quench_plots}, we obtain a singlet-triplet gap of $\omega_{0} = 0.28$ meV, corresponding to a thermal stability of $T_{\text{st}} = 3.2$ K ($T_{\text{st}} = \hbar \omega_{0}/k_B$). Finally, we note that this qubit is universal and can access any point on the Bloch sphere with a specifically chosen quench \cite{Supplementary_material}.
    
    \begin{table}[t]
        \caption{Different GNR dimensions with their corresponding singlet-triplet gaps using the same parameters as in Fig. \ref{Quench_plots}. We give example quenches that yield large amplitude oscillations ($\ge$99\%) alongside their associated Rabi frequency for $\kappa_{\text{R}} = 1$. We consider this case given the lack of sensitivity to changes in the SOC around the high amplitude region. The quench tunability, $\delta\kappa_{z}$, is defined as the range over which the oscillation amplitude is greater than 80\%.}
	\label{GNR_table}
	\begin{ruledtabular}
	\begin{tabular}{cccccc}
		$L \times W$ & $\omega_{0}^{(1)}$ (meV) & $\kappa_{z}$ & $\delta\kappa_{z}$ & $f_{\text{R}}$ (GHz) \\
		\hline
		$36 \times 7$ & 0.28 & 0.8000 & 0.0735 & 24.9 \\ 
            $44 \times 7$ & 1.35 & 0.0975 & 0.0095 & 3.44 \\
            $52 \times 7$ & 1.48 & 0.0116 & 0.0014 & 0.49 \\
            $16 \times 11$ & 0.44 & 0.7050 & 0.0247 & 8.95 \\
            $28 \times 9$ & 1.46 & 0.0287 & 0.0018 & 0.66 \\
            $36 \times 9$ & 1.50 & 6.1 $\times 10^{-4}$ & 5.2 $\times 10^{-5}$ & 0.02 \\
	\end{tabular}
	\end{ruledtabular}
    \end{table}
    
    The dimensions of the GNR also play a central role in determining the characteristics of these 2D nano-qubits. Specifically, $\delta$ decreases rapidly as the GNR size is increased, resulting in a more positive singlet energy, $E_{s} = ( \, \widetilde{U} - \sqrt{4\delta^{2} + \widetilde{U}^{2}} \, )/2$, in larger GNRs, whilst leaving the $T_{-}$ triplet energy, $E_{t} = -\Delta_{\text{xc}}$, unaffected. Naturally, the singlet-triplet gap, and thus thermal stability of the qubit, will also change with the GNR size, see Table \ref{GNR_table}. For GNRs with $E_{t} < E_{s}$, an increase in GNR size will guarantee a larger spin-triplet gap, whilst for GNRs with $E_{s} < E_{t}$, the same occurs with a reduction in GNR size. Regardless, the region of optimal amplitude appears to narrow rapidly as the singlet-triplet gap is increased, whilst still coinciding with smaller Rabi frequencies, thus requiring a higher voltage resolution (required to hone in on $\delta\kappa_{z}$ in Table \ref{GNR_table}) in the experimental apparatus. To explore the large playground of possible GNR dimensions, we analyze how the pre- and post-quench qubit Hamiltonian parameters are affected by changes in length along the armchair and zigzag edges in Ref. \cite{Supplementary_material}. Our extensive numerical studies show that the largest amplitudes correspond to the lowest Rabi frequncies for a given GNR, though, these can still be in excess of 10 GHz. Moreover, the largest GNRs will require more extreme quenches to yield optimal oscillations at the cost of smaller frequencies. Specifically, the Rabi amplitude in these cases becomes $\mathcal{A}_{\text{R}} \simeq (2 g_{\text{R}}^{(2)}/\Delta_{2}^{\null})^{2} = (2 g_{\text{R}}^{(2)}/(hf_{\text{R}}))^{2}$, meaning that high amplitudes will only be achievable by using quenches that dramatically reduce the MEC, see Table \ref{GNR_table}. Interestingly, this interplay of the Rashba mixing term, $g_{\text{R}}$, Rabi frequency, and Rabi amplitude is remarkably reminiscent of the standard Rabi oscillations via periodic driving.

     We briefly comment on the effects that defects in the GNR have upon the dynamics of the nano-qubit. Our numerical studies revealed that even the strongest form of disorder in the form of atomically sharp defects (i.e. vacancies) yielded inconsequential changes to the dynamics of the nano-qubit. We attribute this protection to the topological nature of the edge states \cite{Rizzo_18,Lawrence_20} forming the qubit manifold. Similar robustness against point defects has been previously reported for pseudohelical and helical edge states in proximity-coupled GNRs \cite{Tobias_18,Santos_18}.
    
    
    \textit{\textit{Detection}.---} The detection of this nano-qubit can be achieved using a GNR setup similar to Ref. \cite{Niu2023} with a charge detection scheme based upon a secondary quantum dot as in Ref. \cite{Gachter2022}. Specifically, we suggest using two large tapered graphene flakes placed close together with a small gap between their tips, such that the gap lies on a proximitizing substrate. This gap can be bridged placing a large GNR on top of these flakes, such that the GNR extends far into each flake's region. The GNR section bridging this gap will be the finite GNR that may act as a topological QD with the ability to host a nano-qubit. The graphene flakes can be contacted by standard metallic electrodes to allow for the manipulation of their Fermi levels to enable the loading/unloading of electrons from the GNR. Finally, a perpendicular electric field can be applied to the GNR by using a dual-gate setup. To detect the loading/unloading of electrons from the GNR, we propose that a secondary QD be placed in the vicinity of the GNR, such that the change in charge of the GNR leads to a measurable change in the electrostatic potential experienced by the electrons on the secondary QD. By having this detecting QD tuned in to the steep slope of a Coulomb resonance, these changes in the electrostatic potential will be readily detected by the secondary QD. A schematic of this setup is provided in Ref. \cite{Supplementary_material}. To measure the nano-qubit, we propose following the same approach as Ref. \cite{Gachter2022} but instead of moving the topological QD's energy levels via a plunger gate, the Fermi levels of the graphene flakes should instead be moved simultaneously to emulate the moving the the GNR energy levels. The step-by-step single-shot nano-qubit read-out can be summarised as follows: (i) first load two electrons into the GNR's ground state. (ii) raise the graphene flake Fermi energies above the GNR's excited state and then apply the quench using the dual-gate. (iii) stop the quench and then lower the graphene flake Fermi energies to lie between the ground state and excited state of the GNR, allowing for the excited state to unload if it is occupied followed by the occupation of the ground state.
    
    
    \textit{\textit{Final remarks}.---} Let us briefly discuss the effects of spin relaxation and spin decoherence in our setup. The spin relaxation times reported in chemically synthesised nanoribbons are on the order of milliseconds at $T=10$ K \cite{Slota2018} which is promising. The single-qubit quality factor, $\mathcal{Q} = \pi f_{\text{R}} T_2$ ($T_2$ is the decoherence time), is an important figure of merit for our proposal as it estimates the effectiveness of the 2D nano-qubits in performing a successful quantum computation \cite{DiVincenzo2000}. Our work predicts optimal Rabi oscillations ($P_{\text{max}} \geq 99$\%) and $f_R \sim 100$ GHz in the smallest systems considered \cite{Supplementary_material}. We expect spin-orbit-assisted electron-phonon coupling and hyperfine interactions due to C\textsuperscript{13} isotopes to be two important limiting factors for $T_{2}$, akin to gate-defined graphene QDs \cite{Banzerus_21,Gachter2022}. The prominence of spin-phonon relaxation processes is apparent in the time-resolved ESR measurements in GNRs, dominating the spin dynamics at low temperatures relevant for 2D nano-qubit operation \cite{Slota2018}. We note that the nanoscale nature of these systems provides an intrinsic protection against spatial fluctuations of chemical potential and proximity couplings common in graphene flakes \cite{Locatelli_10,Vicent_17} (i.e. spatial variations in $\lambda_{\text{R}}$ and $\Delta_{\text{xc}}$ due to inhomogeneities in the substrate occur on length scales larger than the GNR), hence reducing decoherence effects originating from the magnetic substrate. The development of a complete microscopic description of these mechanisms in the proximitized GNRs considered here will be an interesting direction for future research. However, we may garner an insight on the values of $T_{2}$ we might expect in these systems, based upon previous studies. With an MEC of 1 meV or larger, reminiscent of a large magnetic field, we anticipate spin-phonon coupling to dominate the spin decoherence rate \cite{Hachiya2014}. For example, taking $T_{2} = 0.4 \, \mu$s from Ref. \cite{Slota2018} yields over $10^{4}$ coherent single-qubit Rabi oscillations ($\mathcal{Q} \sim 10^{4}$) for the fastest nano-qubits with $T_{\text{st}} = 40$ K. Given the fast progress in chemical synthesis of GNRs, we expect near-future systems to achieve even greater quality factors that what we have predicted here.

    In conclusion, we have proposed a type of spin-orbit qubit in graphene-based nanostructures that can be rotated using all-electrical methods, yielding coherent Rabi oscillations with Rabi frequencies in excess of 10 GHz that have thermal stability up to order 10 K. The electrical control proposed here is achieved by harnessing the Rashba SOC and MEC induced by the proximitization of finite graphene nanoribbons. The ability to tune these couplings through the use of an out-of-plane electric field unveils a method for qubit manipulation, a quantum quench, which has the potential to open up avenues in other qubit designs. Finally, we showed that this 2D nano-qubit is universal and can be read out using a simple detection scheme similar to the single-shot read-out method of Elzerman \textit{et al.} \cite{Elzerman2004}. These ultra-fast all-electrical universal nano-qubits are within reach of current bottom-up synthesis methods and offer an alternative route to realising the first graphene-based qubit. The next challenge for these nano-qubits will be the creation of logic gates using GNR arrays on both a theoretical and experimental front. This is likely to require further advancements in bottom-up synthesis, so that atomically precise GNRs with specified dimensions can be produced with good yield and uniformity.
    
    The authors acknowledge support from the Royal Society (London) through Grants No. URF$\backslash$R$\backslash$191021 and RF$\backslash$ERE$\backslash$210281.
    
%

 
      \clearpage
    
    \onecolumngrid
    \section*{Supplementary Material for ``Ultrafast all-electrical universal nano-qubits'}
    \onecolumngrid

    \section*{The Hubbard-Dimer Model}
    
    The full tight-binding Hamiltonian for a proximitized graphene nanoribbon (GNR) with electron-electron interactions can be decomposed as $H = H_{0} + H_{\text{R}} + H_{\text{xc}} + H_{\text{int}}$, where $H_{0}$ describes nearest-neighbor hopping, $H_{\text{R}}$ describes the proximity-induced Rashba spin-orbit coupling (SOC), $H_{\text{xc}}$ describes the magnetic exchange coupling (MEC), and $H_{\text{int}}$ describes electron-electron interactions via the Hubbard interaction. Writing each term out explicitly yields
    \begin{equation}
    \begin{split}
        H = \sum_{\sigma} \sum_{\langle i,j \rangle} t_{ij} c_{i\sigma}^{\dagger} c_{j\sigma}^{\null} + i \frac{2\lambda_{\text{R}}}{3} \sum_{\sigma,\sigma'} \sum_{\langle i,j \rangle} (\mathbf{d}_{ij}^{\null} \times \boldsymbol{\sigma}_{\sigma\sigma'}^{\null})_{z}^{\null} c_{i\sigma}^{\dagger} c_{j\sigma'}^{\null} + \Delta_{\text{xc}} \sum_{\sigma,\sigma'}\sum_{i} \sigma_{z,\sigma\sigma'}^{\null} c_{i\sigma}^{\dagger} c_{i\sigma'}^{\null} +  U \sum_{i} n_{i\uparrow} n_{i\downarrow},
    \end{split}
    \end{equation}
    where $c_{i\sigma}^{(\dagger)}$ are the annihilation (creation) operators for an electron on site $i$ with spin $\sigma$, $n_{\sigma} = c_{i\sigma}^{\dagger} c_{i\sigma}^{\null}$ are the number operators for site $i$ and spin $\sigma$, $\langle i,j \rangle$ denote sums over nearest-neighbors lattice sites, $\mathbf{d}_{ij}$ are the vectors connecting site $j$ to site $i$, $\boldsymbol{\sigma}$ is the vector of Pauli matrices, $\sigma_{\alpha}$ ($\alpha \in \{x,y,z\}$) are Pauli matrices, $t_{ij}$ are the nearest-neighbor hopping energies, $\lambda_{\text{R}}$ is the Rashba SOC strength, and $\Delta_{\text{xc}}$ is the MEC. To obtain the effective Hubbard-Dimer Hamiltonian, we will need to isolate the states corresponding to the quasi-zero energy modes (QZEMs) mentioned in the main text.
    
    The eigenstates of $H_{0}$ may be written as linear superpositions of the individual site states alongside the resolution of identity with the QZEMs isolated,
    \begin{equation}
        \ket{\psi_{n}^{\null}} = \sum_{i = 1}^{N} \psi_{n,i}^{\null} \ket{i}, \qquad \mathcal{I} = \ket{\psi_{a}^{\null}}\bra{\psi_{a}^{\null}} + \ket{\psi_{b}^{\null}}\bra{\psi_{b}^{\null}} + \sum_{n \neq 0^{\pm}} \ket{\psi_{n}^{\null}}\bra{\psi_{n}^{\null}},
    \end{equation}
    where $N$ is the total number of lattice sites, the sum in $\mathcal{I}$ excludes the $\ket{\psi_{0^{\pm}}^{\null}}$ states, and the QZEMs are given by $\ket{\psi_{a}^{\null}} = (\ket{\psi_{0^{+}}^{\null}} + \ket{\psi_{0^{-}}^{\null}})/\sqrt{2}$ and $\ket{\psi_{b}^{\null}} = (\ket{\psi_{0^{+}}^{\null}} - \ket{\psi_{0^{-}}^{\null}})/\sqrt{2}$. By defining a set of creation/annihilation operators in the eigenbasis of $H_{0}$,
    \begin{equation}
        d_{n\sigma}^{\dagger} = \sum_{i = 1}^{N} c_{i\sigma}^{\dagger} \braket{i|\psi_{n\sigma}^{\null}}, \qquad d_{n\sigma}^{\null} = \sum_{i = 1}^{N} c_{i\sigma}^{\null} \braket{\psi_{n\sigma}^{\null}|i}, \qquad c_{i\sigma}^{\dagger} = \sum_{n} d_{n\sigma}^{\dagger} \braket{\psi_{n\sigma}^{\null}|i}, \qquad c_{i\sigma}^{\null} = \sum_{n} d_{n\sigma}^{\null} \braket{i|\psi_{n\sigma}^{\null}},
    \end{equation}
    the simple nearest-neighbor hopping Hamiltonian becomes $H_{0} = \sum_{n,\sigma} \varepsilon_{n} d_{n\sigma}^{\dagger} d_{n\sigma}^{\null}$, where $\varepsilon_{n}$ is the eigenvalue for state $n$. As with the QZEM states, we define the annihilation (creation) operators for the QZEMs as $a_{\sigma}^{(\dagger)} = (d_{0^{+}\sigma}^{(\dagger)} + d_{0^{-}\sigma}^{(\dagger)})/\sqrt{2}$ and $b_{\sigma}^{(\dagger)} = (d_{0^{+}\sigma}^{(\dagger)} - d_{0^{-}\sigma}^{(\dagger)})/\sqrt{2}$. The QZEM component of $H_{0}$ may therefore be isolated as
    \begin{equation}
        H_{0} = \tilde{t} \sum_{\sigma} (a_{\sigma}^{\dagger}b_{\sigma}^{\null} + b_{\sigma}^{\dagger}a_{\sigma}^{\null}) + \sum_{\sigma} \sum_{n \neq 0^{\pm}} d_{n\sigma}^{\dagger} d_{n\sigma}^{\null},
        \label{H0_ab_full_form}
    \end{equation}
    with $\tilde{t} = (\varepsilon_{0^{+}} - \varepsilon_{0^{-}})/2$. Given our focus on the physics around zero energy, we may ignore the third term of Eq. \ref{H0_ab_full_form}.

    Our next step is to handle the Hubbard interaction term, $H_{\text{int}}$. Rewriting this piece in terms of the $d$ operators yields
    \begin{equation}
        H_{\text{int}} = U \sum_{\substack{m,n, \\m',n'}} d_{n\uparrow}^{\dagger} d_{m\uparrow}^{\null} d_{n'\downarrow}^{\dagger} d_{m'\downarrow}^{\null} \sum_{i} \braket{\psi_{n\uparrow}^{\null}|i} \braket{i|\psi_{m\uparrow}^{\null}} \braket{\psi_{n'\downarrow}^{\null}|i} \braket{i|\psi_{m'\downarrow}^{\null}}.
    \end{equation}
    Isolating the the QZEM contribution gives rise to two types of interaction, one acting ``on-site'' and another describing interactions between the QZEMs,
    \begin{equation}
        H_{\text{int},1} = a_{\uparrow}^{\dagger}a_{\uparrow}^{\null} a_{\downarrow}^{\dagger}a_{\downarrow}^{\null} \widetilde{U} + (a \rightarrow b), \qquad H_{\text{int},2} = a_{\uparrow}^{\dagger}a_{\uparrow}^{\null} b_{\downarrow}^{\dagger}b_{\downarrow}^{\null} U \sum_{i} |\braket{\psi_{a,\uparrow}^{\null}|i}|^{2} |\braket{\psi_{b,\downarrow}^{\null}|i}|^{2} + (a \leftrightarrow b),
    \end{equation}
    where $\widetilde{U} = U \eta$, and $\eta = \sum_{i} |\braket{\psi_{a,\uparrow}|i}|^{2} |\braket{\psi_{a,\downarrow}|i}|^{2}$ is the inverse participation ratio. Given that the QZEMs are spatially well separated, the product $|\braket{\psi_{a,\uparrow}^{\null}|i}|^{2} |\braket{\psi_{b,\downarrow}^{\null}|i}|^{2}$ will be small, and hence we may ignore the $H_{\text{int},2}$ contribution. This can be seen by considering that a given site state, $\ket{i}$, must sit entirely on an A site or B site of the graphene layer. For illustrative purposes, let us suppose that $\ket{i}$ lies on an A site. In this case, the overlap of $\ket{i}$ with the $\ket{\psi_{a\sigma}^{\null}}$ QZEM state will be large, meaning that the overlap of the same $\ket{i}$ with the $\ket{\psi_{b\sigma}^{\null}}$ QZEM will be small. In the limit that the QZEMs are completely localized to either the A or B sublattices, this term will vanish. We may therefore approximate the Hubbard interaction around zero energy by $H_{\text{int},1}$.

    Our next port of call is the Rashba SOC. We play the same game as above by taking the expression for $H_{\text{R}}$ and rewriting it in terms of the $d$ operators. For convenience, we define $\lambda_{mn}^{\sigma\sigma'} = i \frac{2}{3} \lambda_{\text{R}}^{\null}\sum_{\langle i,j \rangle} (\mathbf{d}_{ij}^{\null} \times \boldsymbol{\sigma}_{\sigma\sigma'}^{\null})_{z}^{\null} \braket{\psi_{m\sigma}^{\null}|i}\braket{j|\psi_{n\sigma'}^{\null}}$. With this notation, we find the QZEM contribution to $H_{\text{R}}$ to be
    \begin{equation}
    \begin{split}
        \widetilde{H}_{\text{R}} = \frac{1}{2} \sum_{\sigma,\sigma'} \bigg[ (\lambda_{0^{+}0^{+}}^{\sigma\sigma'} &+ \lambda_{0^{+}0^{-}}^{\sigma\sigma'} + \lambda_{0^{-}0^{+}}^{\sigma\sigma'} + \lambda_{0^{-}0^{-}}^{\sigma\sigma'}) a_{\sigma}^{\dagger} a_{\sigma'}^{\null} + (\lambda_{0^{+}0^{+}}^{\sigma\sigma'} + \lambda_{0^{-}0^{-}}^{\sigma\sigma'} - \lambda_{0^{+}0^{-}}^{\sigma\sigma'} - \lambda_{0^{-}0^{+}}^{\sigma\sigma'}) b_{\sigma}^{\dagger} b_{\sigma'}^{\null} \\
        &+ (\lambda_{0^{+}0^{+}}^{\sigma\sigma'} + \lambda_{0^{-}0^{+}}^{\sigma\sigma'} - \lambda_{0^{+}0^{-}}^{\sigma\sigma'} - \lambda_{0^{-}0^{-}}^{\sigma\sigma'}) a_{\sigma}^{\dagger} b_{\sigma'}^{\null} + (\lambda_{0^{+}0^{+}}^{\sigma\sigma'} + \lambda_{0^{+}0^{-}}^{\sigma\sigma'} - \lambda_{0^{-}0^{+}}^{\sigma\sigma'} - \lambda_{0^{-}0^{-}}^{\sigma\sigma'}) b_{\sigma}^{\dagger} a_{\sigma'}^{\null} \bigg].
    \end{split}
    \end{equation}
    This can be simplified greatly by noting that Rashba SOC only permits nearest-neighbor spin-flip hopping. Therefore, the $a^{\dagger}a$ and $b^{\dagger}b$ terms must vanish and $\sigma' = -\sigma = \bar{\sigma}$. These restrictions yield several equations relating the various $\lambda_{mn}^{\sigma\sigma'}$, however, the ones which are of relevance to simplifying the form of the Rashba Hamiltonian are
    \begin{equation}
        \lambda_{0^{+}0^{+}}^{\sigma\bar{\sigma}} + \lambda_{0^{+}0^{-}}^{\sigma\bar{\sigma}} + \lambda_{0^{-}0^{+}}^{\sigma\bar{\sigma}} + \lambda_{0^{-}0^{-}}^{\sigma\bar{\sigma}} = 0, \qquad \lambda_{0^{+}0^{+}}^{\sigma\bar{\sigma}} + \lambda_{0^{-}0^{-}}^{\sigma\bar{\sigma}} - \lambda_{0^{+}0^{-}}^{\sigma\bar{\sigma}} - \lambda_{0^{-}0^{+}}^{\sigma\bar{\sigma}} = 0.
        \label{HR_coeff_restrictions1}
    \end{equation}
    Another restriction can be obtained by recalling that $H_{\text{R}}^{\dagger} = H_{\text{R}}^{\null}$,
    \begin{equation}
        (\lambda_{0^{+}0^{+}}^{\sigma\bar{\sigma}} + \lambda_{0^{-}0^{+}}^{\sigma\bar{\sigma}} - \lambda_{0^{+}0^{-}}^{\sigma\bar{\sigma}} - \lambda_{0^{-}0^{-}}^{\sigma\bar{\sigma}})^{*} = \lambda_{0^{+}0^{+}}^{\bar{\sigma}\sigma} + \lambda_{0^{+}0^{-}}^{\bar{\sigma}\sigma} - \lambda_{0^{-}0^{+}}^{\bar{\sigma}\sigma} - \lambda_{0^{-}0^{-}}^{\bar{\sigma}\sigma}.
        \label{HR_coeff_restrictions2}
    \end{equation}
    Combining Eq. \ref{HR_coeff_restrictions1} and Eq. \ref{HR_coeff_restrictions2} yields
    \begin{equation}
        (\lambda_{0^{+}0^{+}}^{\sigma\bar{\sigma}} - \lambda_{0^{+}0^{-}}^{\sigma\bar{\sigma}})^{*} = \lambda_{0^{+}0^{-}}^{\bar{\sigma}\sigma} + \lambda_{0^{+}0^{-}}^{\bar{\sigma}\sigma} = \tilde{\lambda}_{\text{R},\sigma}^{*},
    \end{equation}
    and hence
    \begin{equation}
        \widetilde{H}_{\text{R}} = \sum_{\sigma} \left[ \tilde{\lambda}_{\text{R},\sigma}^{\null} a_{\sigma}^{\dagger} b_{\bar{\sigma}}^{\null} + \tilde{\lambda}_{\text{R},\sigma}^{*} b_{\bar{\sigma}}^{\dagger} a_{\sigma}^{\null} \right].
    \end{equation}
    When constructing the effective $6\times6$ Hamiltonian at the end of this section, we will make use of $\tilde{\lambda}_{\text{R},\sigma} = \text{sgn}(\sigma) \tilde{\lambda}_{\text{R}}$.

    Finally, we turn our attention to the MEC part of the Hamiltonian. In the $H_{0}$ eigenbasis the MEC Hamiltonian becomes
    \begin{equation}
        H_{\text{xc}} = \Delta_{\text{xc}} \sum_{\sigma,\sigma'} \sum_{n} (d_{n\uparrow}^{\dagger} d_{n\uparrow}^{\null} - d_{n\downarrow}^{\dagger} d_{n\downarrow}^{\null})
    \end{equation}
    Focusing on the QZEMs, we obtain
    \begin{equation}
        \widetilde{H}_{\text{xc}} = \Delta_{\text{xc}} \sum_{\sigma,\sigma'} \sum_{n} \Big[ a_{\uparrow}^{\dagger}a_{\uparrow}^{\null} + b_{\uparrow}^{\dagger}b_{\uparrow}^{\null} + a_{\downarrow}^{\dagger}a_{\downarrow}^{\null} + b_{\downarrow}^{\dagger}b_{\downarrow}^{\null} \Big].
    \end{equation}

    Therefore, with the isolation of the QZEM contributions to each piece of the Hamiltonian, we write the effective QZEM Hamiltonian as
    \begin{equation}
        \widetilde{H} = \sum_{\sigma} (\tilde{t} \, a_{\sigma}^{\dagger}b_{\sigma}^{\null} + \tilde{\lambda}_{\text{R},\sigma} \, a_{\sigma}^{\dagger}b_{\bar{\sigma}}^{\null} + h.c.) + \sum_{\nu = a,b} \Big[ \Delta_{\text{xc}} (\nu_{\uparrow}^{\dagger}\nu_{\uparrow}^{\null}-\nu_{\downarrow}^{\dagger}\nu_{\downarrow}^{\null}) + \widetilde{U}(\nu_{\uparrow}^{\dagger}\nu_{\uparrow}^{\null} \nu_{\downarrow}^{\dagger}\nu_{\downarrow}^{\null}) \Big].
        \label{Hubbard_Dimer_Hamiltonian_SM}
    \end{equation}
    Regarding the computation of the parameters appearing above, we note that the effective hopping and Rashba SOC energies may be calculated by taking the relevant matrix elements of the original Hamiltonian pieces with respect to the the QZEM states, $\ket{\psi_{a\sigma}^{\null}}$. That is to say, $\tilde{t} = \bra{\psi_{a\sigma}^{\null}} H_{0} \ket{\psi_{b\sigma}^{\null}}$ and $\tilde{\lambda}_{\text{R},\sigma} = \bra{\psi_{a\sigma}^{\null}} H_{\text{R}} \ket{\psi_{b\bar{\sigma}}^{\null}}$. In the case of the Hubbard interaction, this relies upon the computing the overlap of the QZEMs with all of the original site states to determine the inverse participation ratio.
    
    By projecting this effective Hubbard-Dimer model onto the two-particle basis $\{\ket{2,0},\ket{0,2},\ket{\uparrow,\uparrow},\ket{\downarrow,\downarrow},\ket{\downarrow,\uparrow},\ket{\uparrow,\downarrow}\}$, we obtain the following $6\times6$ Hamiltonian,
    \begin{equation}
        \widetilde{H} = \begin{pmatrix}
            \widetilde{U} & 0 &  -\tilde{\lambda}_{\text{R}}^{\null} & -\tilde{\lambda}_{\text{R}}^{\null} & -\tilde{t} & \tilde{t} \\
            0 & \widetilde{U} &  -\tilde{\lambda}_{\text{R}}^{*} & -\tilde{\lambda}_{\text{R}}^{*} & -\tilde{t} & \tilde{t} \\
            -\tilde{\lambda}_{\text{R}}^{*} & -\tilde{\lambda}_{\text{R}}^{\null} &  \Delta_{\text{xc}} & 0 & 0 & 0 \\
            -\tilde{\lambda}_{\text{R}}^{*} & -\tilde{\lambda}_{\text{R}}^{\null} &  0 & -\Delta_{\text{xc}} & 0 & 0 \\
            -\tilde{t} & -\tilde{t} &  0 & 0 & 0 & 0 \\
            \tilde{t} & \tilde{t} &  0 & 0 & 0 & 0
        \end{pmatrix}.
    \end{equation}
    
    \vspace{0.25cm}
    
    \section*{The Qubit Hamiltonian}

    To construct the effective qubit Hamiltonian, let us include the Rashba SOC perturbatively. The eigenvalues of $\widetilde{H}$ with $\lambda_{\text{R}} = 0$ are readily found to be
    \begin{equation}
        E_{1},E_{2} \in \left\{\frac{1}{2} \left( \widetilde{U} - \sqrt{16 \tilde{t}^{2} + \widetilde{U}^{2}} \right), \, -\Delta_{\text{xc}}\right\}, \quad E_{3} = 0, \quad E_{4} = \Delta_{\text{xc}}, \quad E_{5} = \widetilde{U}, \quad E_{6} = \frac{1}{2} \left( \widetilde{U} + \sqrt{16 \tilde{t}^{2} + \widetilde{U}^{2}} \right).
    \end{equation}
    Given the size of $\widetilde{U}$, the states associated to $E_{5}$ and $E_{6}$ are well gapped out from the first four energy levels. We identify these remaining four states according to their $z$-component of spin expectation value. Clearly the states with energies 0 and $\pm\Delta_{\text{xc}}$ are spin triplets that are degenerate in the absence of the Zeeman splitting yielded by the exchange coupling, with $\ket{T_{0}}$ associated with $E_{3}$ and $\ket{T_{\pm}}$ associated with $E = \pm \Delta_{\text{xc}}$. Naturally, this leaves the state with $E = \left( \widetilde{U} - \sqrt{16 \tilde{t}^{2} + \widetilde{U}^{2}} \right)/2$ as the singlet state, $\ket{S}$. As illustrated in Fig. 1b of the main text, the qubit will be formed of $\ket{S}$ and $\ket{T_{-}}$. We therefore define the qubit gap as
    \begin{equation}
        \omega_{0} = -\Delta_{\text{xc}} + \frac{1}{2} \left( \sqrt{16 \tilde{t}^{2} + \widetilde{U}^{2}} - \widetilde{U} \right),
    \end{equation}
    which will enter the effective qubit Hamiltonian, $\mathcal{H}$, via the term $-\omega_{0} \sigma_{z}$ written in the basis $\{\ket{S},\ket{T_{-}}\}$. We account for the Rashba SOC to leading order by calculating the matrix elements $\bra{i} \widetilde{H}_{\text{R}} \ket{j}$ ($\ket{i,j} \in \{\ket{S},\ket{T_{-}}\}$), where $\widetilde{H}_{\text{R}}$ is the Hamiltonian piece containing only the Rashba SOC. We find that $\bra{i} \widetilde{H}_{\text{R}} \ket{j} = g_{\text{R}} (1 - \delta_{ij})$, with
    \begin{equation}
        g_{\text{R}} = \frac{8 \, \tilde{t} \, \widetilde{\lambda}_{\text{R}}}{\widetilde{U} + \sqrt{16 \, \tilde{t}^{2} + \widetilde{U}^{2}}} \left( 2 + \frac{32 \, \tilde{t}^{2}}{\left( \widetilde{U} + \sqrt{16 \, \tilde{t}^{2} + \widetilde{U}^{2}} \right)^{2}} \right)^{-1/2}.
    \end{equation}
    We may therefore write the effective qubit Hamiltonian for the GNR as
    \begin{equation}
	\mathcal{H} = -\frac{\omega_{0}}{2} \sigma_{z} + g_{\text{R}} \sigma_{x},
    \end{equation}
    where $\sigma_{i}$ are Pauli matrices.
    
    \vspace{0.25cm}
    
    \subsection*{Qubit Rotation: The Quantum Quench}
    
    To determine the effects of a sudden change in $\Delta_{\text{xc}}$ and $\lambda_{\text{R}}$, we will need to understand how the initial state of the system and how it subsequently evolves with time. The eigenvectors of this effective qubit Hamiltonian are easily found to be
    \begin{equation}
	\ket{\chi_{\pm}} = \frac{1}{\sqrt{E_{\pm}(2 E_{\pm}-\omega_{0})}} \begin{pmatrix}
		E_{\pm} - \frac{\omega_{0}}{2} \\
		g_{\text{R}}
	\end{pmatrix}
    \end{equation}
    with eigenvalues $E_{\pm} = \pm \frac{1}{2} \sqrt{4 g_{\text{R}}^{2} + \omega_{0}^{2}}$. Given that $\ket{\chi_{\pm}}$ are simple linear combinations of $\ket{S}$ and $\ket{T_{-}}$, we may decompose them as
    \begin{equation}
	\ket{\chi_{-}^{\null}} = \cos\psi \ket{S} + \sin\psi \ket{T_{-}}, \qquad \ket{\chi_{+}^{\null}} = \sin\psi \ket{S} - \cos\psi \ket{T_{-}},
	\label{Trig_states}
    \end{equation}
    without loss of generality.

    Let us take the system to have the parameters $\omega_{0}^{(1)}$ and $g_{\text{R}}^{(1)}$ prior to the quench. Applying a quantum quench will shift the parameters $\omega_{0}^{(1)} \rightarrow \omega_{0}^{(2)}$ and $g_{\text{R}}^{(1)} \rightarrow g_{\text{R}}^{(2)}$. The evolution of the system is then determined using the qubit Hamiltonian with these post-quench parameters, $\mathcal{H}_{2}$. After a time $t$, a system initialised in the pre-quench ground state, $\ket{\chi_{-}^{(1)}}$, will find itself in the time-evolving state $\ket{\psi(t)} = e^{-i\mathcal{H}_{2}t} \ket{\chi_{-}^{(1)}}$. Consequently, the probability of finding our system in the pre-quench excited state is given by $P(t) = |\braket{\chi_{+}^{(1)}|\psi(t)}|^{2}$. The overlap of the pre-quench excited state with the system's state can be shown to be
    \begin{equation}
        \braket{\chi_{+}^{(1)}|\psi(t)} = i e^{-iE_{-}^{(2)}t} e^{-i \frac{\Delta_{2}}{2} t} \sin\left( \frac{\Delta_{2}}{2} t \right) \sin(2\psi_{12}).
    \end{equation}
    where $\psi_{12} = \psi_{1}-\psi_{2}$, the subscripts on $\psi_{i}$ denote their pre- (1) and post- (2) quench values, and $\Delta_{2} = E_{+}^{(2)}  - E_{-}^{(2)}$ is the difference in eigenvalues for $\mathcal{H}_{2}$. This allows us to quickly obtain the probability of finding the system in the original excited state as
    \begin{equation}
	P(t) = \sin^{2}\left( \frac{\Delta_{2}}{2} t \right) \sin^{2}(2\psi_{12}),
    \end{equation}
    with the singlet-triplet mixing simplifying to
    \begin{equation}
        \sin^{2}(2\psi_{12}) = \frac{4 \left(g_{\text{R}}^{(2)} \omega_{0}^{(1)}-g_{\text{R}}^{(1)} \omega_{0}^{(2)} \right)^2}{\left(4 g_{\text{R}}^{(1)2}+\omega_{0}^{(1)2} \right) \left(4 g_{\text{R}}^{(2)2}+\omega_{0}^{(2)2}\right)}.
        \label{ST_mixing_explicit}
    \end{equation}
    For completeness, we find the overlap of the pre-quench ground state and time-evolving system state to be
    \begin{equation}
        \braket{\chi_{-}^{(1)}|\psi(t)} = e^{-iE_{-}^{(2)}t} e^{-i \frac{\Delta_{2}}{2} t} \left[ \cos\left( \frac{\Delta_{2}}{2} t \right) + i \sin\left( \frac{\Delta_{2}}{2} t \right) \cos(2\psi_{12}) \right].
    \end{equation}
    
    \begin{figure}
        \centering
        \includegraphics[width=0.9\textwidth]{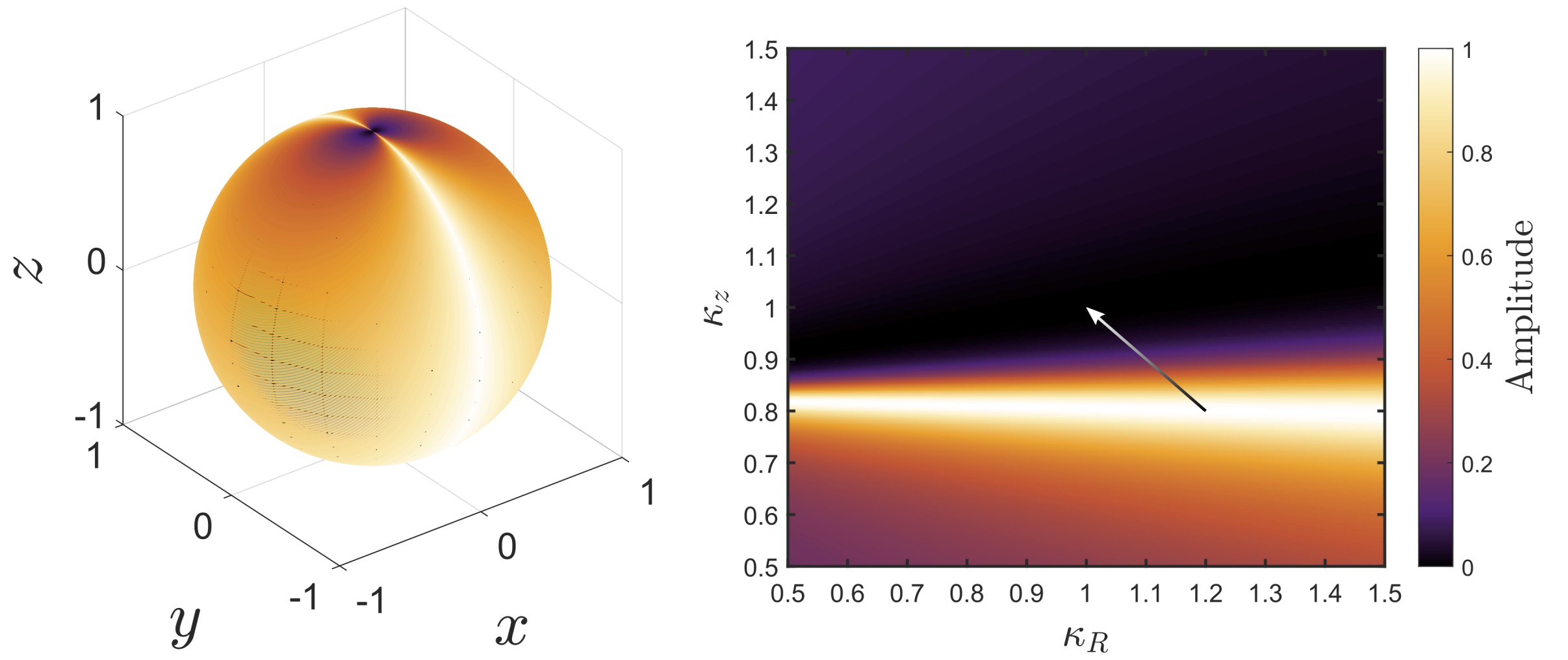}
        \caption{Left: Qubit rotations around the Bloch sphere for various quenches. Right: Rabi oscillation amplitude quench dependence for the GNR used in Fig. 2 of the main text. The line connects from an optimal quench giving Rabi oscillations of 100\% to a null quench (i.e. no change). The qubit orbits (left) are taken from quenches along this line.}
        \label{Bloch_sphere}
    \end{figure}
    
    \vspace{0.25cm}
    
    \section*{Universality of the Nano-Qubit}
    
    Now that we have acquired the probability of transition between the excited and ground states of the original system due to the quantum quench, we now address the question ``how much of the Bloch sphere is covered by the qubit?'' We may write the general state of the qubit may be written as
    \begin{equation}
        \ket{\psi(t)} = \cos\left(\frac{\theta}{2}\right) e^{-i\phi} e^{i\gamma} \ket{\chi_{-}^{(1)}} + \sin\left(\frac{\theta}{2}\right) e^{i\gamma} \ket{\chi_{+}^{(1)}},
	\label{qubit}
    \end{equation}
    where $\theta$ and $\phi$ are the angles lying on the Bloch sphere ($\theta \in [0,\pi]$ and $\phi \in [0,2\pi]$), and $e^{i\gamma}$ is an arbitrary phase factor. For a qubit to be universal, the angles $\theta$ and $\phi$ should be able to explore the whole Bloch sphere (i.e. the qubit can access the whole space of $(\theta,\phi)$ coordinates). Note that this choice of parameterisation locates $\ket{\chi_{-}^{\null}}$ at the top of the Bloch sphere and $\ket{\chi_{+}^{\null}}$ at the bottom of the Bloch sphere.
    
    To obtain the ranges that $\theta$ and $\phi$ may explore as permitted by the physical parameters, we note that eq. \ref{qubit} implies
    \begin{equation}
    \begin{split}
        \cos\left(\frac{\theta}{2}\right) e^{-i\phi} e^{i\gamma} &= e^{-iE_{-}^{(2)}t} e^{-i \frac{\Delta_{2}}{2} t} \left[ \cos\left( \frac{\Delta_{2}}{2} t \right) + i \sin\left( \frac{\Delta_{2}}{2} t \right) \cos(2\psi_{12}) \right], \\
	\sin\left(\frac{\theta}{2}\right) e^{i\gamma} &= i e^{-iE_{-}^{(2)}t} e^{-i \frac{\Delta_{2}}{2} t} \sin\left( \frac{\Delta_{2}}{2} t \right) \sin(2\psi_{12}).
    \end{split}
    \end{equation}
    By identifying $e^{i\gamma} = i e^{-iE_{-}^{(2)}t} e^{-i \frac{\Delta_{2}}{2} t}$, these relations simplify to
    \begin{equation}
    \begin{split}
	\cos\left(\frac{\theta}{2}\right) e^{-i\phi} &= \sin\left( \frac{\Delta_{2}}{2} t \right) \cos(2\psi_{12}) - i \cos\left( \frac{\Delta_{2}}{2} t \right), \\
	\sin\left(\frac{\theta}{2}\right) &= \sin\left( \frac{\Delta_{2}}{2} t \right) \sin(2\psi_{12}).
    \end{split}
    \end{equation}
    From this, we may immediately write down the equations governing the qubit's position on the Bloch sphere,
    \begin{equation}
        \tan\phi = \cot\left( \frac{\Delta_{2}}{2} t \right) \sec(2\psi_{12}), \qquad \sin\left(\frac{\theta}{2}\right) = \sin\left( \frac{\Delta_{2}}{2} t \right) \sin(2\psi_{12}).
    \end{equation}
    We note that the relation giving $\theta$ may permit negative solutions. However, we can map these negative solutions onto the range $\theta \in [0,\pi]$ by taking their absolute value (i.e. $\theta \rightarrow |\theta|$ if $\theta < 0$) if we also shift $\phi$ by $\pm \pi$. We may do this given that $\tan\phi$ has a period of $\pi$ and hence $\phi$ and $\phi + n \pi$ ($n \in \mathbb{Z}$) are valid solutions. The universality of this qubit can be seen numerically in Fig. \ref{Bloch_sphere}, where we show the orbits traversed around the Bloch sphere by the qubit over time for different quenches over the full range of Rabi oscillation amplitudes.
    
    \vspace{0.25cm}
    
    \section*{Experimental Setup}
    
    The detection of this nano-qubit requires the isolation of a proximitized GNR whose length is on the order of tens of atoms. In order to do this, we propose a setup similar to that of Niu et al. \cite{Niu2023}, wherein a long GNR is used to bridge the gap between two large graphene sheets. This gap can be tuned to be the size of the GNRs discussed here, which in turn allows us to create a nano-qubit in the GNR region bridging this gap. The GNR bridge system can be created using the bottom-up methods of Ref. \cite{Niu2023} on top of some substrate. A small trench may also be created in this substrate to host a proximitizing substrate, such as a transition metal dichalcogenide to introduce strong SOC. Furthermore, if two proximitizing layers are needed, an additional material may be placed on top of the GNR.
    
\begin{figure}
        \centering
        \includegraphics[width=\linewidth]{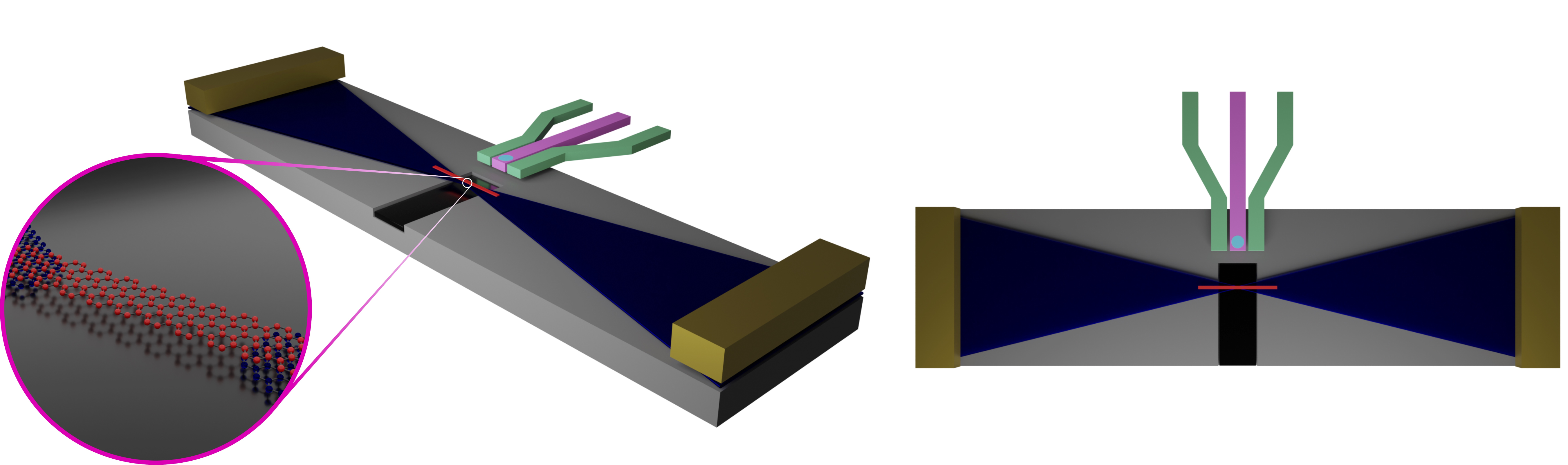}
        \caption{Sketch of an experimental setup to detect the nano-qubit hosted by a GNR. The red strip represents a typical GNR that is extended in one dimension. The blue triangles represent the large graphene sheets placed in contact with the GNR, with a small gap between their tips. This gap is bridged by the GNR and its size is on the order of the finite GNRs discussed here required to create a nano-qubit. The large graphene sheets are contacted by metal electrodes (gold blocks). The large grey region denotes a substrate upon which the graphene is placed or grown. A trench can be introduced beneath the GNR bridge in which a proximitizing subtrate can be placed, such as a transition metal dichalcogenide, as well as a back-gate. Finally, a QD charge detector (pale blue dot) can be created using a finger gate (pink block) placed between two electron reservoirs (pale green blocks). Left: three-dimensional perspective of a possible experimental setup with a zoom-in on the GNR bridge. Right: Top-down view of the same setup.}
        \label{Setup_schematic}
    \end{figure}

    In order to drive Rabi oscillations via the quantum quench protocol, we will also need to make use of an out-of-plane electric field. A back-gate may also be placed in the substrate trench to enable this; a dual-gate set up may also be used. Additionally, we propose that the loading/unloading of electrons from the GNR can be achieved by shifting the Fermi levels of the graphene sheets. This can be achieved by using metallic electrodes placed in contact with the graphene sheets. Finally, single-shot readout of the GNR can be achieved using a QD charge detector similar to that of G\"{a}chter et al. \cite{Gachter2022}. Such a detector is created by using a finger gate to create a QD that is tuned into a Coulomb resonance peak. The change in the potential on this sensing QD due to a single electron hopping onto the GNR will then lead sharp changes in the voltage measured across the QD, thus allowing for the detection of quantum transport. In Fig. \ref{Setup_schematic} we present the sketch of a possible experimental setup that makes use of the features discussed above in order to create, control, and readout the nano-qubits we propose to be hosted in GNRs.
    
    \vspace{0.25cm}

    \begin{table}[t]
        \caption{Different GNR dimensions with their corresponding singlet-triplet gaps using the same parameters as in Fig. 2 of the main text. We give example quenches that yield large amplitude oscillations ($\ge$99\%) alongside their associated Rabi frequency for $\kappa_{\text{R}} = 1$. We consider this case given the lack of sensitivity to changes in the SOC around the high amplitude region. The quench tunability, $\delta\kappa_{z}$, is defined as the range over which the oscillation amplitude is greater than 80\%.}
	\label{GNR_table}
	\begin{ruledtabular}
	\begin{tabular}{cccccc}
		$L \times W$ & $\omega_{0}^{(1)}$ (meV) & $\kappa_{z}$ & $\delta\kappa_{z}$ & $f_{\text{R}}$ (GHz) \\
		\hline
		$36 \times 7$ & 0.28 & 0.8000 & 0.0735 & 24.9 \\ 
            $44 \times 7$ & 1.35 & 0.0975 & 0.0095 & 3.44 \\
            $52 \times 7$ & 1.48 & 0.0116 & 0.0014 & 0.49 \\
            $16 \times 11$ & 0.44 & 0.7050 & 0.0247 & 8.95 \\
            $28 \times 9$ & 1.46 & 0.0287 & 0.0018 & 0.66 \\
            $36 \times 9$ & 1.50 & 6.1 $\times 10^{-4}$ & 5.2 $\times 10^{-5}$ & 0.02 \\
	\end{tabular}
	\end{ruledtabular}
    \end{table}
    
    \section*{Quench Protocol for Various Graphene Nanoribbons}

    In this section we present the effects of the quantum quench upon a variety of different GNRs to illustrate the range of possible nano-qubits available. We will see that the dimensions of the GNR play a central role in determining the characteristics of these 2D nano-qubits. Specifically, $\delta$ decreases rapidly as the GNR size is increased, resulting in a more positive singlet energy, $E_{s} = ( \, \widetilde{U} - \sqrt{4\delta^{2} + \widetilde{U}^{2}} \, )/2$, in larger GNRs, whilst leaving the $T_{-}$ triplet energy, $E_{t} = -\Delta_{\text{xc}}$, unaffected. Naturally, the singlet-triplet gap, and thus thermal stability of the qubit, will also change with the GNR size, see Table \ref{GNR_table}. For GNRs with $E_{t} < E_{s}$, an increase in GNR size will guarantee a larger spin-triplet gap, whilst for GNRs with $E_{s} < E_{t}$, the same occurs with a reduction in GNR size. The region of optimal amplitude appears to narrow rapidly as the GNR size is increased, whilst still coinciding with smaller Rabi frequencies, thus requiring a higher voltage resolution (required to hone in on $\delta\kappa_{z}$ in Table \ref{GNR_table}) in the experimental apparatus.

\begin{figure}[t]
        \centering
        \includegraphics[width=0.8\linewidth]{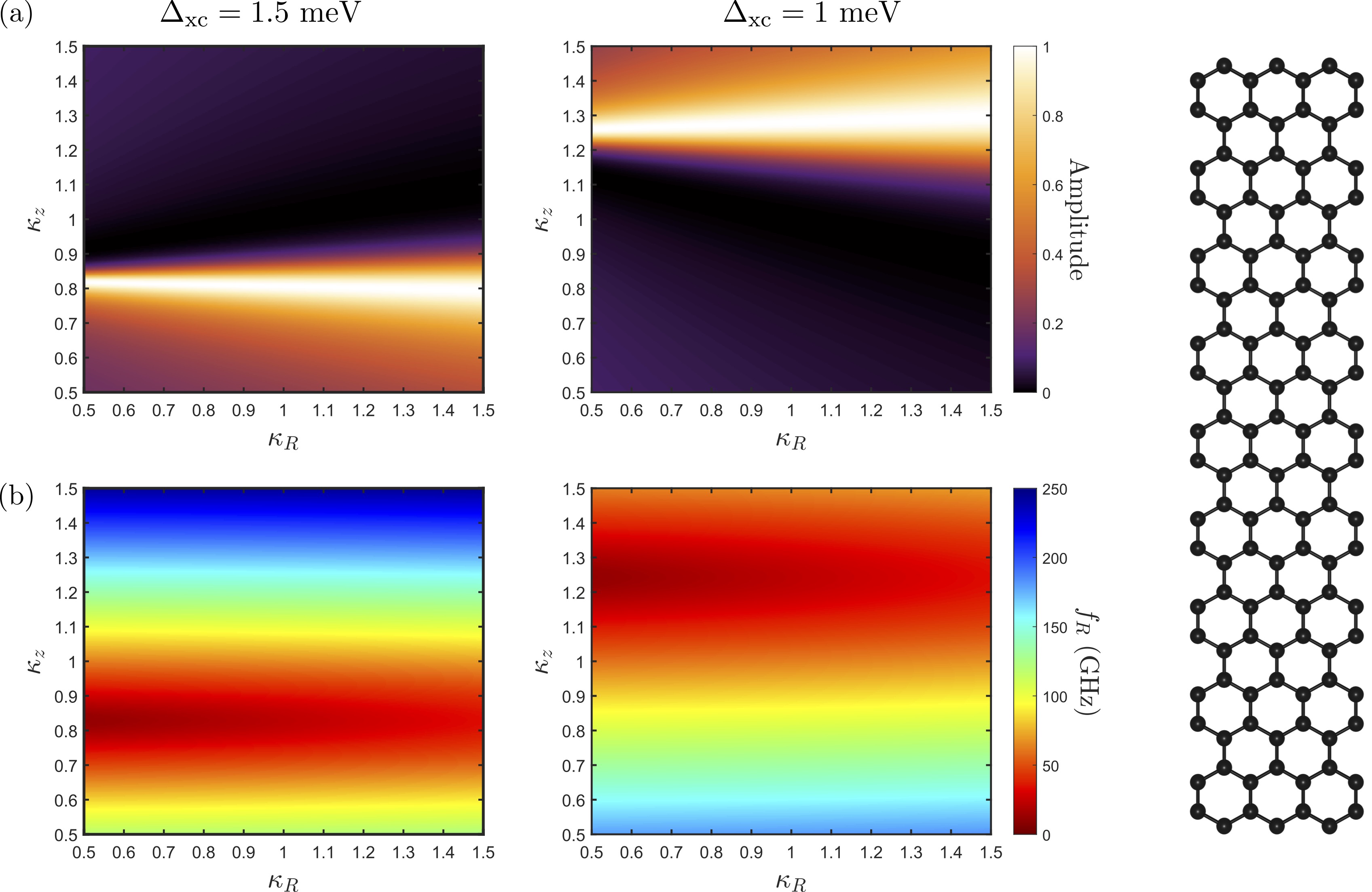}
        \caption{Rabi oscillation amplitudes (a) and frequencies (b) for a $36 \times 7$ atom GNR (right) with different MEC values. Both GNRs use $\lambda_{\text{R}}^{(1)} = 10$ meV, $t = 2.7$ eV, Hubbard interaction $U = 2t$. The MECs used are $\Delta_{\text{xc}}^{(1)} = 1.5$ meV (left) and $\Delta_{\text{xc}} = 1$ meV (right). For $\Delta_{\text{xc}} = 1$ meV, a quench of $\kappa_{\text{R}} = 1.2$ and $\kappa_{z} = 1.29$ yields an amplitude of 99.9\% with $f_{\text{R}} = 29.8$ GHz.}
        \label{MEC_quench_comparison_plots}
    \end{figure}

    To probe how the parameters and GNR dimensions affect the nano-qubit, let us start by probing the effect of the MEC upon the optimal quench region in Fig. \ref{MEC_quench_comparison_plots}. Here we use the same GNR as in Fig. 2 of the main text ($36 \times 7$ atom GNR) but with different values for $\Delta_{\text{xc}}$. Specifically, the left plots show the effects of a quench upon a GNR whose qubit manifold's ground state is the $\ket{T_{-}}$ state, whilst the right plots consider a GNR with the $\ket{S}$ state as its qubit ground state instead. We can see that the optimal quench region corresponds to reducing the post-quench gap, $\Delta_{2}$ (see Fig. 1b of the main text). This can be easily understood from Eq. \ref{ST_mixing_explicit}, whose denominator is directly proportional to $\Delta_{2}$, which naturally leads to the largest oscillation amplitudes as $\Delta_{2} \rightarrow 0$. Note that the gap can never vanish whilst Rashba coupling is present though ($g_{\text{R}} \propto \lambda_{\text{R}}$). Again, we see that the Rabi oscillations are largely insensitive to the change in Rashba SOC, which is a consequence of how $\omega_{0}$ and $g_{\text{R}}$ depend upon $\tilde{\lambda}_{\text{R}}$ and $\Delta_{\text{xc}}$. Specifically, we can choose a quench such that $\omega_{0}$ is reduced greatly due to it being a difference in the zero-Rashba singlet and triplet energies, which in turn yields a drastic change in $\Delta_{2}$. In contrast, most attainable quenches will only alter $g_{\text{R}}$ by at most 50\% and thus not change $\Delta_{2}$ greatly.
    
\begin{figure}
        \centering
        \includegraphics[width=0.7\linewidth]{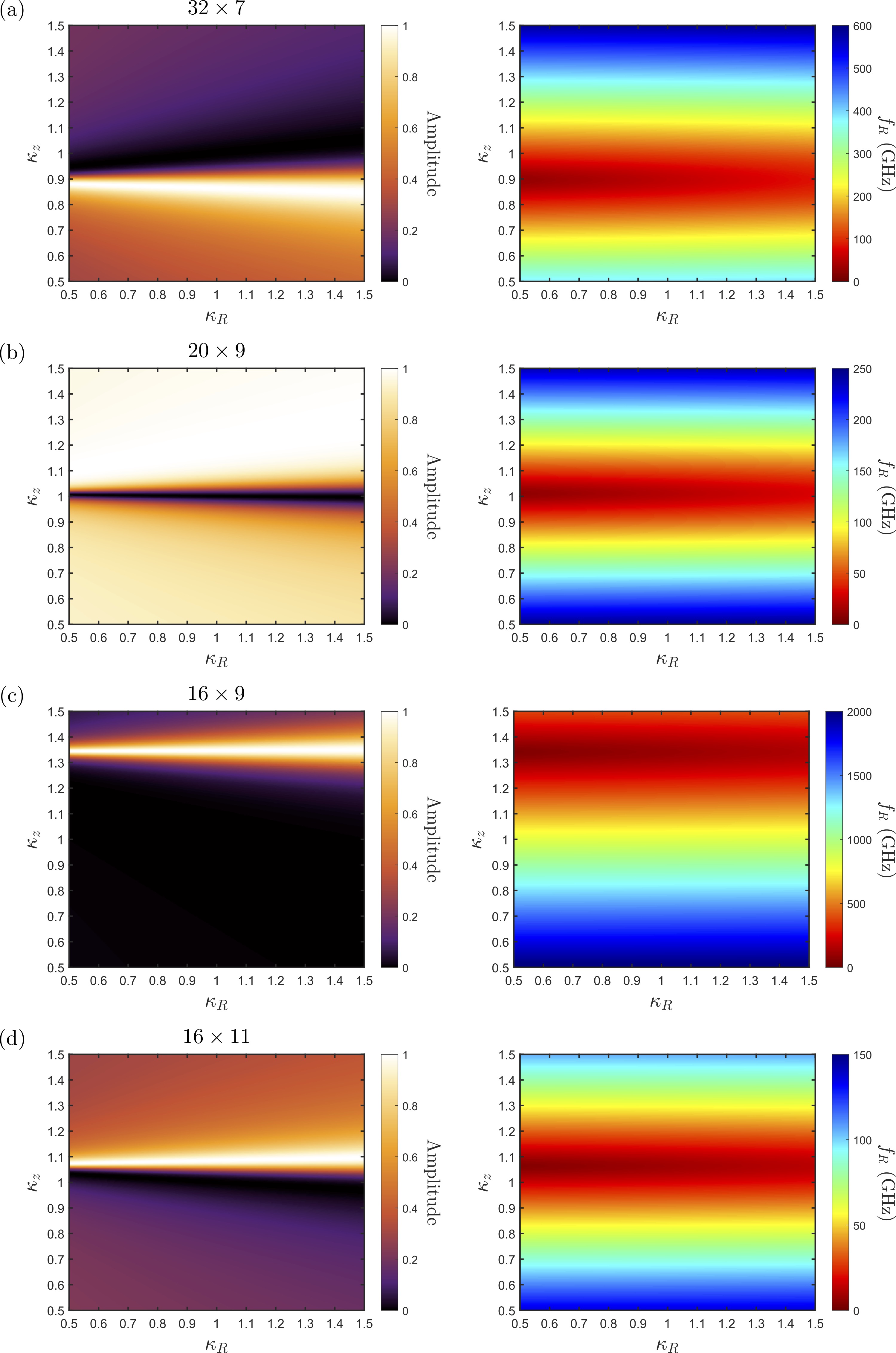}
        \caption{Quench dependence for $32 \times 7$ (a), $20 \times 9$ (b), $16 \times 9$ (c), and $16 \times 11$ atom GNRs. All cases use $\lambda_{\text{R}} = 10$ meV, $t = 2.7$ eV, Hubbard interaction $U = 2t$. The MEC is taken to be $\Delta_{\text{xc}} = 4$ meV (a), 2 meV (b), 10 meV (c), and 1 meV (d). Examples of optimal quenches are as follows: (a) $\kappa_{\text{R}} = 0.87$ and $\kappa_{z} = 0.84$ yields an amplitude of 99.9\% with $f_{\text{R}} = 53.6$ GHz, (b) $\kappa_{\text{R}} = 1.2$ and $\kappa_{z} = 1.06$ yields an amplitude of 100\% with $f_{\text{R}} = 94.4$ GHz, (c) $\kappa_{\text{R}} = 1.16$ and $\kappa_{z} = 1.35$ yields an amplitude of 99.6\% with $f_{\text{R}} = 120.78$ GHz, (d) $\kappa_{\text{R}} = 1.06$ and $\kappa_{z} = 1.09$ yields an amplitude of 99.3\% with $f_{\text{R}} = 11.5$ GHz. The quench tunability for each of these GNRs is $\delta\kappa_{z} = 0.0833$ (a), 0.0431 (c), 0.0541 (d), whilst their qubit stability temperatures are $T_{\text{st}} = 5.57$ K (a), 1.02 K (b), 39.81 K (c), 0.84 K (d). The quench tunability of the $20 \times 9$ GNR cannot be defined and is discussed in the main supplementary text.}
        \label{Quench_plots_long_GNRs}
    \end{figure}

    The nano-qubits so far discussed in $36 \times 7$ atom GNRs can also be observed in a wide range of other GNR dimensions. In Fig. \ref{Quench_plots_long_GNRs}, we present the effects of the quantum quench upon $32 \times 7$, $20 \times 9$, $16 \times 9$, and $16 \times 11$ atom GNRs. In all cases, as above, we see that the optimal quench region corresponds to reducing $\Delta_{2}$. Furthermore, the region of the largest Rabi osicllations continues to coincide with the region of lowest Rabi frequencies. The reason for this can be determined from Eq. \ref{ST_mixing_explicit} in combination with $f_{\text{R}} \propto \Delta_{2}$. Given our previous discussion of larger Rabi oscillations prefering a reduction in the post-quench gap, we can see this is immediately counter-productive for the Rabi frequency, which prefers larger post-quench gaps. However, whilst the oscillation amplitude and frequency might behave in opposite manners regarding the quantum quench, the frequencies observed in the ``low'' frequency region can still be on the order of 100 GHz. Of particular note is the $20 \times 9$ GNR, which exhibits a very large set of possible quenches that generate Rabi oscillations with near-perfect amplitudes. In fact, for a fixed Rashba coupling, this system will only exhibit Rabi oscillations with amplitudes less than 80\% over a very narrow region, $0.83 \leq \kappa_{z} \leq 1.06$. Amplitudes of at least 90\% can be obtained for any $\kappa_{z} < 0.53$ or $\kappa_{z} > 1.08$ for a fixed Rashba coupling. We note that the MEC quench required to reach these large amplitudes will need to be larger for increases in the Rashba SOC, but smaller for a decrease in the Rashba SOC, see Fig. \ref{Quench_plots_long_GNRs}b.
    
\begin{figure}
        \centering
        \includegraphics[width=0.7\linewidth]{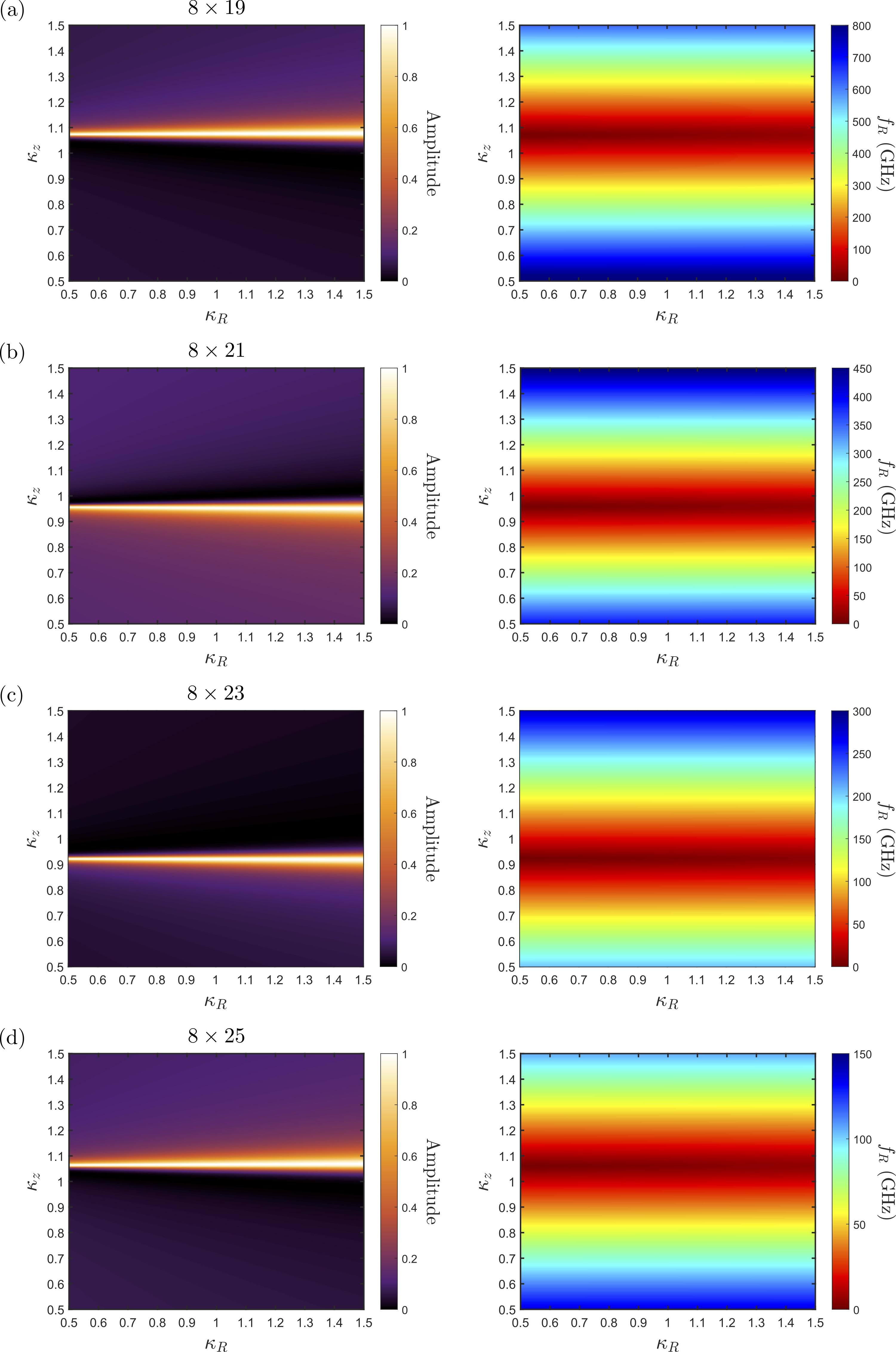}
        \caption{Quench dependence for $8 \times 19$ (a), $8 \times 21$ (b), $8 \times 23$ (c), and $8 \times 25$ (d) atom GNRs. All cases use $\lambda_{\text{R}} = 10$ meV, $t = 2.7$ eV, Hubbard interaction $U = 2t$. The MEC is taken to be $\Delta_{\text{xc}} = 6$ meV (a), 3.5 meV (b), 2 meV (c), and 1 meV (d). Examples of optimal quenches are as follows: (a) $\kappa_{\text{R}} = 1.05$ and $\kappa_{z} = 1.074$ yields an amplitude of 99.9\% with $f_{\text{R}} = 27$ GHz, (b) $\kappa_{\text{R}} = 1.05$ and $\kappa_{z} = 0.95$ yields an amplitude of 99.2\% with $f_{\text{R}} = 15.5$ GHz, (c) $\kappa_{\text{R}} = 1.05$ and $\kappa_{z} = 0.92$ yields an amplitude of 99.1\% with $f_{\text{R}} = 7.83$ GHz, (d) $\kappa_{\text{R}} = 1.05$ and $\kappa_{z} = 1.066$ yields an amplitude of 99.96\% with $f_{\text{R}} = 4.74$ GHz. The quench tunability for each of these GNRs is $\delta\kappa_{z} = 0.0187$ (a), 0.0186 (b), (c) 0.0161, and 0.0198 (d), whilst their qubit stability temperatures are $T_{\text{st}} = 5.03$ K (a), 1.83 K (b), (c) 1.86 K, and 0.73 K (d).}
        \label{Quench_plots_wide_GNRs2}
    \end{figure}

\begin{figure}
        \centering
        \includegraphics[width=0.7\linewidth]{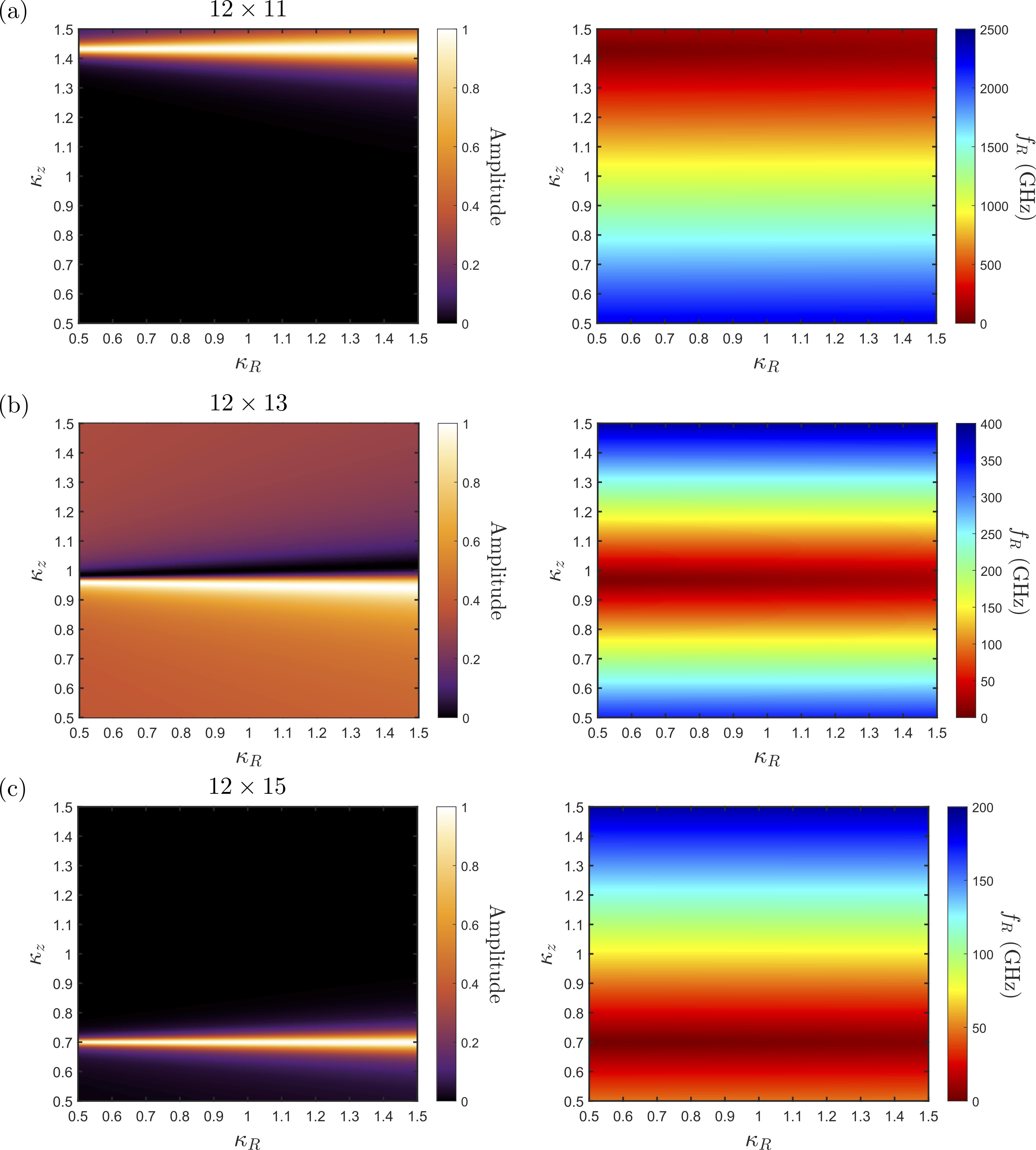}
        \caption{Quench dependence for $12 \times 11$ (a), $12 \times 13$ (b), and $12 \times 15$ (c) atom GNRs. All cases use $\lambda_{\text{R}} = 10$ meV, $t = 2.7$ eV, Hubbard interaction $U = 2t$. The MEC is taken to be $\Delta_{\text{xc}} = 10$ meV (a), 3 meV (b), and 1 meV (c). Examples of optimal quenches are as follows: (a) $\kappa_{\text{R}} = 1.23$ and $\kappa_{z} = 1.43$ yields an amplitude of 99.7\% with $f_{\text{R}} = 101.1$ GHz, (b) $\kappa_{\text{R}} = 1.05$ and $\kappa_{z} = 0.95$ yields an amplitude of 99.7\% with $f_{\text{R}} = 22.2$ GHz, (c) $\kappa_{\text{R}} = 1.1$ and $\kappa_{z} = 0.7$ yields an amplitude of 99\% with $f_{\text{R}} = 4.85$ GHz. The quench tunability for each of these GNRs is $\delta\kappa_{z} = 0.0342$ (a), 0.0436 (b), and 0.0183 (c), whilst their qubit stability temperatures are $T_{\text{st}} = 49.95$ K (a), 1.44 K (b), and 3.5 K (c).}
        \label{Quench_plots_wide_GNRs}
    \end{figure}

    For a fixed width, reducing the length of the GNR results in an increase in the energy gap between the in-gap states, $\delta$, and hence an increase in the QZEM hybridisation energy, $\tilde{t} = \delta/2$. This therefore leads to a lower singlet energy and hence requires a larger MEC in order to place the triplet energy level sufficiently close to the singlet. Likewise, the magnitude of $\tilde{\lambda}_{\text{R}}$ also increases with a reduction in GNR length, however, due to its scale (0.1 meV to 1 meV), the effect of changes here upon the energy levels are negligible in comparison to the changes in $\delta$. In contrast, $\eta$, and hence $\widetilde{U}$, remain largely unaffected by changes in the GNR length, exhibiting only small shifts in their values. Naturally, GNRs whose armchair edge is too short will have ground state singlet energies that are gapped out from the rest of the two-particle spectrum. To balance the reduction in GNR length, we can increase the width (zigzag edge) of the GNR to prevent the singlet energy from lowering too far. For example, a $12 \times 11$ atom GNR has a singlet energy of around -14 meV, but using a $12 \times 13$ atom GNR raises this to around -3 meV. However, both $\delta$ and $\tilde{\lambda}_{\text{R}}$ are much more sensitive to changes in the zigzag edge length than the armchair edge length, thus admitting greater changes in the qubit manifold energies due to changes in the zigzag edge. Figs. \ref{Quench_plots_wide_GNRs2} and \ref{Quench_plots_wide_GNRs} show the effects of a quench for different GNRs with reduced armchair edges and varying zigzag edge lengths. One general trend that can be seen in comparing Fig. \ref{Quench_plots_long_GNRs} to Figs. \ref{Quench_plots_wide_GNRs2} and \ref{Quench_plots_wide_GNRs} is the preference for GNRs with notably longer armchair edges. The region of large Rabi amplitudes for GNRs whose armchair edge is notably longer than their zigzag edge appears to be typically wider in $\kappa_{z}$. This means that honing in on these nano-qubits is likely to be easier in GNRs with long armchair edges and short zigzag edges.
    
 \end{document}